\documentclass[aps,prl,preprint,superscriptaddress]{revtex4-1}

\usepackage{graphicx}
\usepackage[colorlinks=true, urlcolor=blue, pdfborder={0 0 0}]{hyperref}

\usepackage[left]{lineno}

\usepackage{amsmath}
\usepackage{subfigure}
\usepackage{epsfig} 
\usepackage{bm}
\usepackage{epstopdf}
\usepackage{float} 
\usepackage{tikz}
\usepackage[percent]{overpic}
\usepackage{comment}
\usepackage{placeins}
\usepackage{multirow}
\usepackage{rotating}
\usepackage{hhline}
\usepackage{xcolor}
\newcommand{\Arrow}[1][]
{
\begin{tikzpicture}
\draw[<->, line width=0.5mm] (0,0) -- (0.7,0);
\end{tikzpicture}
}

\begin{document}

\title{Coherence-Induced Quantum Forces}


\author{Tarek A. Elsayed}
\email{tarek.ahmed.elsayed@gmail.com}
\affiliation{Department of Physics, School of Science and Engineering, The American University in Cairo, AUC Avenue, P.O. Box 74, New Cairo, 11835, Egypt}


\date{\today}

\begin{abstract}
We introduce a model that explains the phenomenon of correlation-assisted tunneling and puts it in a broader context. This model assumes the existence of an effective force of pure quantum nature between nearby fragments of correlated matter that results due to interference effects. The magnitude of this force depends on the amount of coherence between different locations; it attains a maximum value for fragments in a perfect superfluid state and disappears entirely when the fragments are in the Mott Insulator state. The force can also be explained in terms of the Bohmian quantum potential. We illustrate the implications of this force on the transport of cold atoms through simple potential structures, the  triple-well harmonic trap and optical lattices.

\end{abstract}

\maketitle

\section{Introduction}
Quantum coherence effects play a very important role in the onset of the non-classical states of matter such as superfluidity. Furthermore, the existence of coherence in a quantum system leads to  interesting phenomena in its dynamics and physical properties \cite{streltsov2017,chatterjee2019}. For example, coherence is crucial in inducing the quantum interference effects in Josephson tunneling \cite{jaklevic1964} which has important practical applications. 
Recently, a lot of evidence has been shown  for the interplay between coherence and the transport of matter and energy  excitations in other systems such as biological systems \cite{huelga2013} and cold atom systems \cite{cao2012}. In particular, it has been noticed that the existence of coherence can induce tunneling across potential barriers.   In this regard, coherence provides a new resource for controlling the tunneling dynamics of quantum matter. 

The physics of quantum tunneling is so rich that the interdependence of tunneling and many physical effects has led to new phenomena such as
coherence assisted tunneling \cite{cao2012}, chaos assisted tunneling \cite{arnal2020}, photon-assisted tunneling \cite{sias2008} and phonon-assisted tunneling \cite{oberli1990}. In this article, we put coherence assisted tunneling on a more solid ground and study it in the context of the quantum forces affecting the motion of particles in one dimension (1D). In doing that, we provide a different perspective on coherence assisted tunneling, examine it in the  context of generic transport of quantum matter across potential boundaries and relate it to the quantum interference phenomenon. Our main result is a conceptual one, namely, correlation gives rise to effective non-classical forces that can be detected in cold atoms experiments. The existence of effective forces of pure quantum nature such as Casimir force \cite{casimir1948} and the anti-centrifugal \cite{cirone2001} forces has been noticed long ago. Similar to Casimir forces, the effective force we introduce here depends on the wave nature of the particles. We present several examples to illustrate this effect in the transport of bosonic cold atoms across potential boundaries and periodic structures.

The coherence between two separate locations in a many-body system means the existence of a first-order correlation between those locations. In a superfluid (SF) state of matter, such as the Bose-Einstein Condensate (BEC), there is a perfect phase correlation between any pair of locations since all the particles  occupy the same one-particle quantum state. On the other hand, in a Mott insulator (MI) state, there is a complete absence of first-order correlation between different locations of localized fragments. The contrast between the two regimes leads to several interesting effects such as the dependence of the tunneling of quantum particles in multi-well traps or optical lattices on the amount of coherence between the lattice sites.  It was shown, based on ab initio calculation,  that correlation induced tunneling can be used to probe the amount of coherence between different wells in an optical lattice or the condensed fraction in a BEC \cite{elsayed2018,cao2012}. The explanation we give for this effect is based on the  interference between different fragments of  quantum matter due to the phase coherence. The phase coherence will lead to either constructive or destructive interference depending on the sign of the spatial correlation, giving rise to an effective attractive or repulsive force between the two fragments. We shall also see that this force can be explained in terms of the gradient of the one-particle Bohm's quantum potential. 

\section{Theory}
Let us illustrate this phenomenon by a simple example (henceforth, we use the words coherence and correlations interchangeably). Consider two overlapping fragments of quantum matter in 1D composed of non-interacting particles  localized at $\pm r_0$. In one case, the fragments exist in a perfect superfluid state, $\Psi_{\text{SF}}$, with all the atoms occupying the same orbital  $\phi_0(x)$ (the condensate order parameter), while in another case, the atoms occupy the less correlated, fully-symmetric Fock state $\Psi_{\text{MI}}\equiv |nn\rangle$, which means that $n$ atoms occupy one orbital $\phi_+(x)$ and $n$ atoms occupy another one $\phi_-(x)$ (we take $n=1$, since the one-particle properties of interest in this work will not depend on the total number of particles).  The two orbitals are orthonormal and can be defined in terms of two other orbitals, $\phi_L(x)$ and $\phi_R(x)$ as \(\phi_+(x)=\lambda_+(\phi_L(x)+\phi_R(x)) \) and \(\phi_-(x)=\lambda_-(\phi_L(x)-\phi_R(x)) \) where $\lambda_+$ and $\lambda_-$ are normalization constants and $\phi_L(x)$ and $\phi_R(x)$ are gaussian orbitals centered at $\pm r_0$ and correspond to the left and right fragments. For a many-body system described by the wavefunction $\Psi(x_1,x_2,\dots x_N)$, the one-body reduced density matrix (RDM) is defined as $\rho^{(1)}(x_1,x_1')=\int \Psi(x_1,x_2,\dots x_N)\Psi^*(x_1',x_2,\dots x_N)dx_2\dots dx_N$.
The single-particle density function $\rho(x)$ is given by the diagonal elements of $ \rho^{(1)}$. The coherence between two different locations $x$ and $x'$  is described by the first order correlation function $g^{(1)}(x,x')$ which depends on the off-diagonal elements  of the reduced one-body density matrix and is defined as  \(g^{(1)}(x_1,x_1')=\frac{\rho^{(1)}(x_1,x_1')}{\sqrt{\rho(x_1)\rho(x_1')}}\)  \cite{note}.

We consider three examples of freely evolving systems with distinct correlations, each composed of two fragments symmetrically localized around the origin. One system is described by the MI state, $\Psi_{\text{MI}}$, and two systems by the SF state. The two SF states are $\Psi_{\text{SF}_1}$, whose single particle orbital,  $\phi_0(x)$, gives rise to the same $\rho(x)$ of $\Psi_{\text{MI}}$, i.e., $\phi_0(x)=\sqrt{\rho(x)_{\text{MI}}}$ and $\Psi_{\text{SF}_2}$ whose $\phi_0(x)=\phi_-(x)$. We take $\phi_{L/R}=\sqrt{\frac{\pi}{2}}e^{-(x\pm r_0)^2}$  with $r_0=1$ (atomic units are used throughout this article, setting $\hbar=m=1$). The two $\phi_0$ orbitals are shown in Fig. 1i. We show in Figs. 1a-1c the correlation functions for all the three states, $\Psi_{\text{SF}_1},\Psi_{\text{MI}}$ and $ \Psi_{\text{SF}_2}$ respectively. We notice the existence of a strong correlation between the left and right sides for $\Psi_{\text{SF}_1}$, an anti-correlation for $\Psi_{\text{SF}_2}$ due to the opposite phases of the wavefunction and almost an absence of off-diagonal correlation for $\Psi_{\text{MI}}$. In the middle region, the squeezed off-diagonal correlation in $\Psi_{\text{MI}}$ is controlled by the overlap between the left and right orbitals. The coherence (anti-coherence) exhibited by $\Psi_{\text{SF}_1}$ ($\Psi_{\text{SF}_2}$)  will lead to constructive (destructive) interference when the state evolves with time, creating more (less) populated region in the middle region compared to the uncorrelated Mott state.

\begin{figure*}[t!] \setlength{\unitlength}{0.1cm}
\begin{picture}(120 , 110 )
{

\put(-7, 75){\includegraphics[width=4.5 cm]{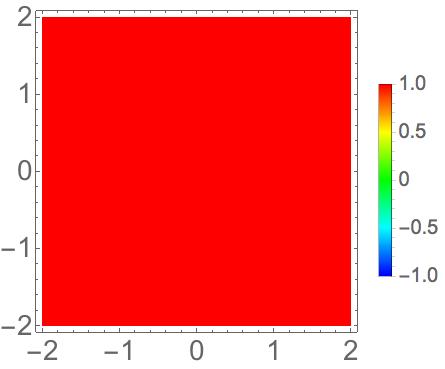}}
\put(45, 75){\includegraphics[width=4.5 cm]{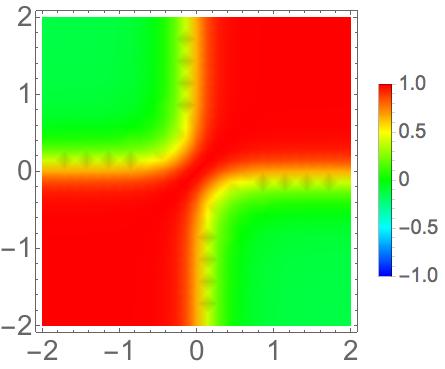}}
\put(98, 75){\includegraphics[width=4.5 cm]{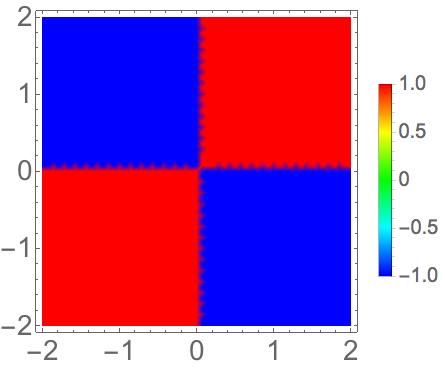}}
\put(91, 34){\includegraphics[width=5.5cm]{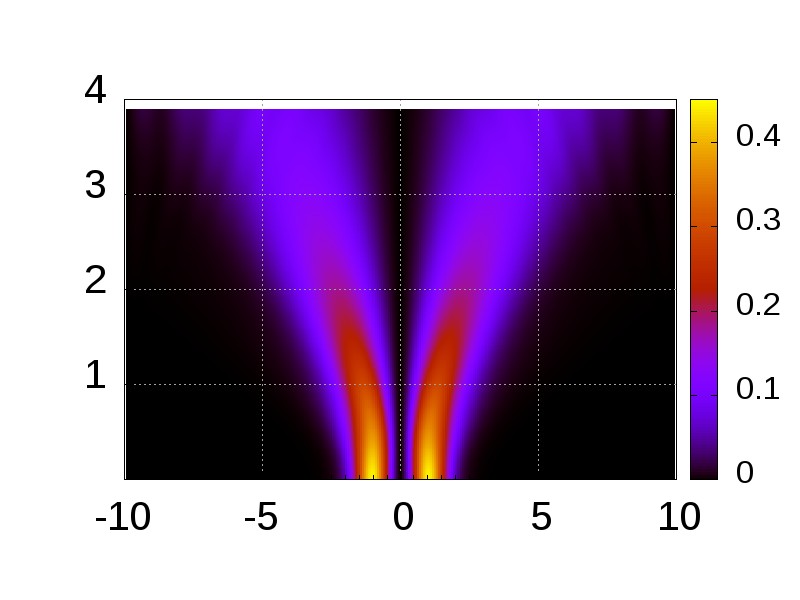}}
\put(38, 34){\includegraphics[width=5.5cm]{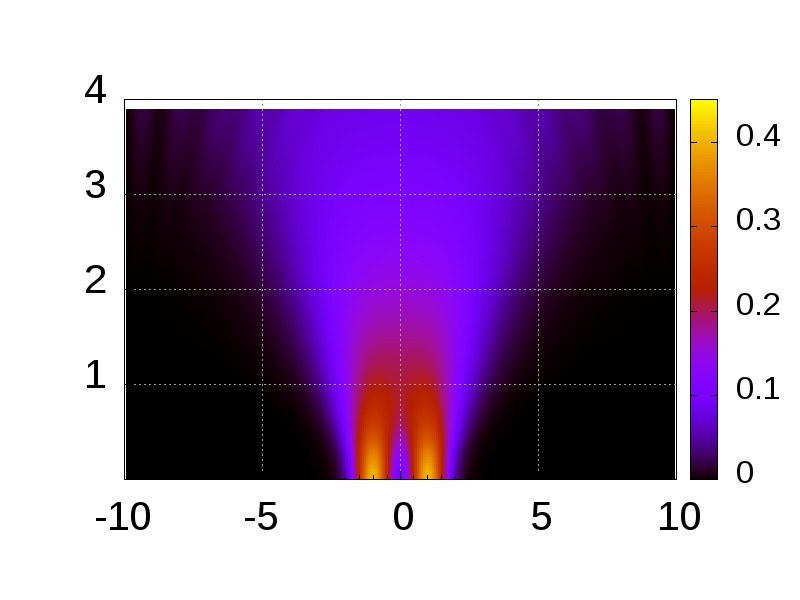}}
\put(-15, 34){\includegraphics[width=5.5cm]{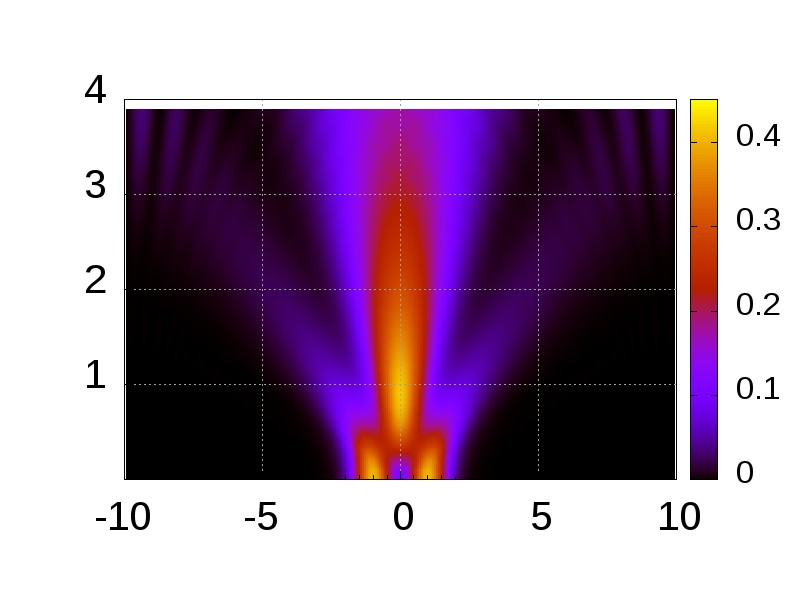}}

\put(-10, 0){\includegraphics[width=5 cm]{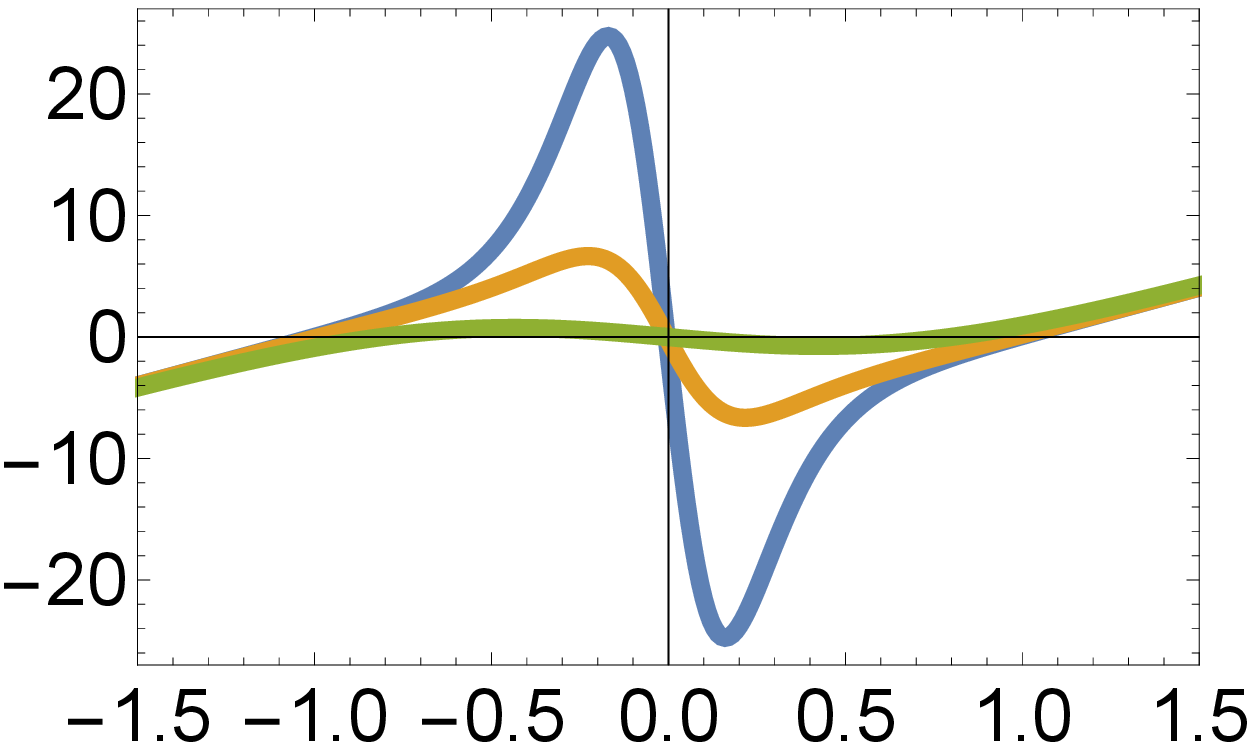}}
\put(47, 0){\includegraphics[width=4.5 cm]{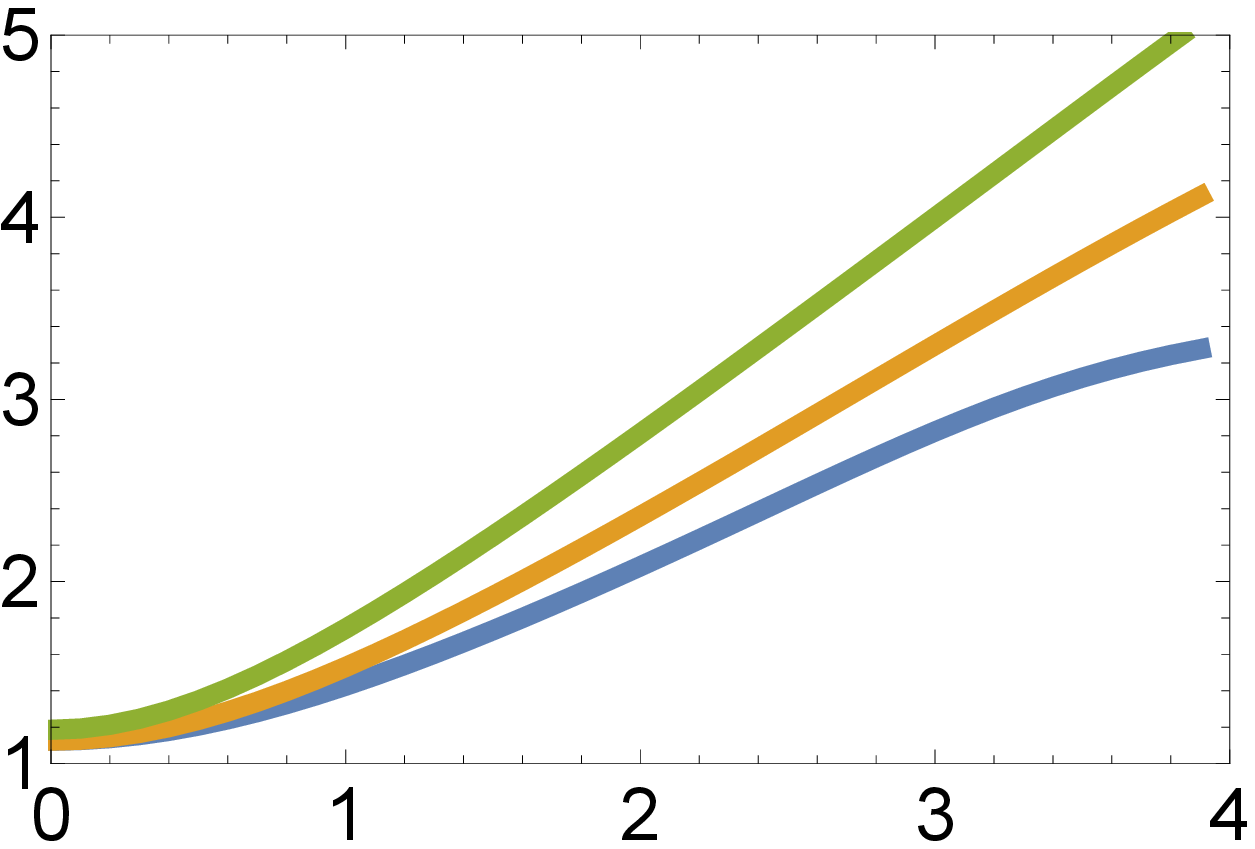}}
\put(98, 0){\includegraphics[width=4.5 cm]{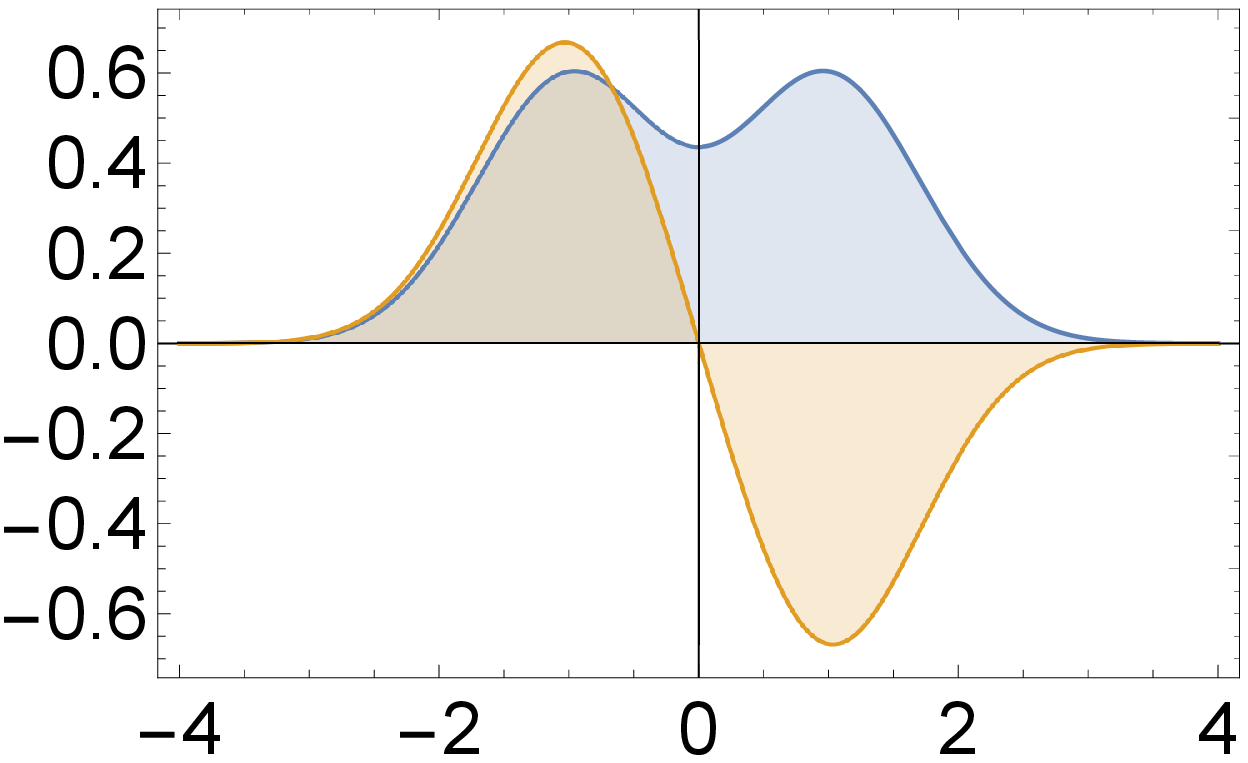}}

\scriptsize

\put(-10, 112){\textbf{a}}
\put(42, 112){\textbf{b}}
\put(94, 112){\textbf{c}}

\put(-10, 73){\textbf{d}}
\put(42, 73){\textbf{e}}
\put(94, 73){\textbf{f}}

\put(-10, 31){\textbf{g}}
\put(42, 31){\textbf{h}}
\put(97, 31){\textbf{i}}

\put(13, -3) {$x$ (m) }
\put(67, -3) {$t$ (s) }

\put(-13, 13) {\rotatebox{90} {$F_q$ (N)} }
\put(42, 11) {\rotatebox{90} {$\Delta x^2$ (m$^2$)} }

\put(8, 73) {$x_1$ (m) }
\put(-10, 90) {\rotatebox{90} {$x_1'$ (m)}}
\put(42, 90) {\rotatebox{90} {$x_1'$ (m)}}
\put(95, 90) {\rotatebox{90} {$x_1'$ (m)}}

\put(60, 73) {$x_1$ (m) }
\put(114, 73) {$x_1$ (m) }

\put(62, 35) {space (m) }
\put(-14, 50) {\rotatebox{90} {Time (s)}}

\put(118, -3){$x$ (m)}
\put(95, 12) {\rotatebox{90} {$\phi_0(x)$ } }

}
\end{picture}
\caption{\textbf{Freely propagating fragments of quantum matter with different correlations.}\\ \textbf{a-c} The first order correlation function $g^{(1)}(x_1,x_1')$ for the  states $\Psi_{\text{SF}_1}$, $\Psi_{\text{MI}}$ and $\Psi_{\text{SF}_2}$ respectively. \textbf{d-f} The space-time density plots for the quantum matter initialized in $\Psi_{\text{SF}_1}$, $\Psi_{\text{MI}}$ and $\Psi_{\text{SF}_2}$ respectively  during their free propagation. \textbf{g} The Bohmian force $F_q$ per particle computed for $\Psi_{\text{SF}_1}$ (blue), $\Psi_{\text{MI}}$ (orange) and $\Psi_{\text{SF}_2}$ (green) in the middle region. \textbf{h} The average single-particle position variance $\Delta x^2$ during the time evolution of $\Psi_{\text{SF}_1}$ (blue), $\Psi_{\text{MI}}$ (orange) and $\Psi_{\text{SF}_2}$ (green). \textbf{i} The initial orbitals $\phi_0(x)$ for $\Psi_{\text{SF}_1}$ (blue) and $\Psi_{\text{SF}_2}$ (orange). Atomic units are used throughout this article, setting $\hbar=m=1$.  }
\label{fig1} 
\end{figure*}

Effectively, the distinction between the dynamics of the two correlated states and the uncorrelated state can be thought of as due to extra attractive (repulsive) forces that cause the fragments to come closer (further apart) from each other as shown in Figs. 1d-1f. We can characterize the effect of this force by the growth of the position variance $\Delta x^2$ which for a single particle system is defined as $\Delta x^2=\langle (x -\langle x \rangle)^2\rangle$. The variance is a global property computed over the whole space and due to the symmetry of our problem we have $\langle x_i \rangle=0$ and we can simply use $\Delta x^2= \int x^2 \rho(x) dx$. Fig. 1h illustrates the growth of $\Delta x^2$ at different rates for the three states, qualitatively consistent with the effective force model introduced above; fragments attracted towards each other more strongly will have smaller $\Delta x^2$ than fragments moving towards $x=0$ less strongly. The dynamics of $\Psi_{\text{MI}}$ represents an average behavior between the symmetric $\Psi_{\text{SF}_1}$ and anti-symmetric $\Psi_{\text{SF}_2}$.

 The numerical simulation of the dynamics presented in Fig. 1 were obtained by evolving each of the three states using the  Multi-Configurational Time-Dependent Hartree method for Bosons (MCTDHB)\cite{streltsov2007,alon2008,lode2012,lode2020,lin2020,mctdh-b}. MCTDHB uses an ansatz for the many-particle wavefunction as a superposition of all possible configurations of symmetrized Hartree product states representing $N$ particles occupying $M$ orbitals. Since there is no interaction between the particles, we use one single orbital for the the SF states, which makes the simulation equivalent to a one-particle problem,  and two orbitals for the MI states. The behavior of the later case resembles the behavior of a single-particle system in a mixed state. We use rigid boundary conditions, but the short time scale of the dynamics ensures that reflections off the boundaries don't severely affect the presented behavior. MCTDHB is also used to simulate the quantum dynamics in the rest of this article. Although it seems unorthodox to talk about forces between different parts of the same wavefunction of a single particle, the effect reported here is manifest in the behavior of non-interacting many-particle systems, whose average behavior can be represented by the wavefunction of single particle systems.

Now, let us shed some light on this phenomenon with the help of the de-Broglie Bohm interpretation of quantum mechanics \cite{bohm1952-I,bohm1952-II}. This interpretation has gained some popularity recently as a tool for calculation and as a resource for gaining new insights into quantum experiments \cite{oriols2007,gondran2005,elsayed2018entangled}.  According to this interpretation, quantum particles follow deterministic trajectories in the physical space under the effect of the Bohmain potential $Q(x)$ which encapsulates all the  quantum effects of the system \cite{bohm1952-I}. Also, for a perfect superfluid, the gradient of the phase of the order parameter plays the role of a velocity field for the Bohmian trajectories. Generally, for a many-body system whose density function in the configuration space $\rho(x_1,\dots x_N)=|\Psi(x_1,\dots x_N)|^2$, the quantum potential is defined as $Q(x_1,\dots x_N)=\frac{-\hbar^2}{2m}\frac{\nabla^2 R(x_1,\dots x_N)}{R(x_1,\dots x_N)}$, where $R(x_1,\dots x_N)=\sqrt{\rho(x_1,\dots x_N)}$. The average one-body Bohm potential $q(x)$ should satisfy $\int q(x_1)\rho(x_1)dx_1=   \int Q(x_1,\dots x_N) \rho(x_1,\dots x_N)dx_1 \dots dx_N$, therefore, $q(x_1)=\int \frac{Q(x_1,\dots x_N)\rho(x_1,\dots x_N) dx_2 \dots dx_N}{\rho(x_1)} $. The Bohmian quantum force per particle can then be defined as $F_q(x)=-\nabla q(x)$. For single particle states such as $\phi_L(x)$ or $\phi_R(x)$, the quantum force will depend only on the probability density and will lead to the dispersion of the wavepacket under the effect the quantum pressure of $\rho(x)$. For many-particle states, more subtle effects will contribute to $F_q(x)$.

We show in Fig. 1g the Bohmian quantum force per particle computed in the middle region for the three initial quantum states. We notice a greater attractive force between  the two fragments in  $\Psi_{\text{SF}_1}$ compared to $\Psi_{\text{MI}}$ and a diminished force in $\Psi_{\text{SF}_2}$ compared to $\Psi_{\text{MI}}$.  We emphasize here that the distinction between the behavior of the quantum force per particle for  $\Psi_{\text{SF}_1}$ and $\Psi_{\text{MI}}$ in the middle region is due to the different coherence behavior in the two cases, not the different quantum pressure of the probability density, since they have the same single particle density by construction. In other words, the quantum force per particle for a correlated system consists of two components: the uncorrelated part (the orange plot) and the correlated part, which can add to or subtract from the former one.

It is clear that the distinction outlined above is most prominent in the initial behavior, before correlations build up between the intermediate region and the outer regions, even for the initially uncorrelated state. We conjecture that this force will ensue in the transient dynamics between any two nearby fragments of coherent quantum matter.  We further anticipate that this effect can be best observed in trapped quantum matter, or at the boundaries between regions of different potential strengths. We illustrate this point in Fig. 2 where we compare  between the transport of the non-interacting particles described by $\Psi_{\text{SF}_1}$ and $\Psi_{\text{MI}}$ in three different cases: transport through a square quantum well (Fig. 2a), transport through a potential step (Fig. 2b) and transport through a square potential barrier whose height is twice the energy per particle (Fig. 2c).  Since the fragments are initially almost  non-overlapping, we simply take $\phi_0(x)=\phi_+(x)$ for  $\Psi_{\text{SF}_1}$. In Figs. 2a and 2c, the initial orbitals $\phi_{L/R}$ are localized at $x=\pm 3$  while the boundaries of the well/barrier are located at $x=\pm 1$. The step potential in Fig. 2b is located at $x=0$ while the initial orbitals are localized at $x=\pm 2$. The height of the barriers in all the three cases is twice as much as the energy per particle.



\begin{figure*}[t!] \setlength{\unitlength}{0.1cm}
\begin{picture}(120 , 95 )
{
\put(-10, 70){\includegraphics[width=4.5 cm ]{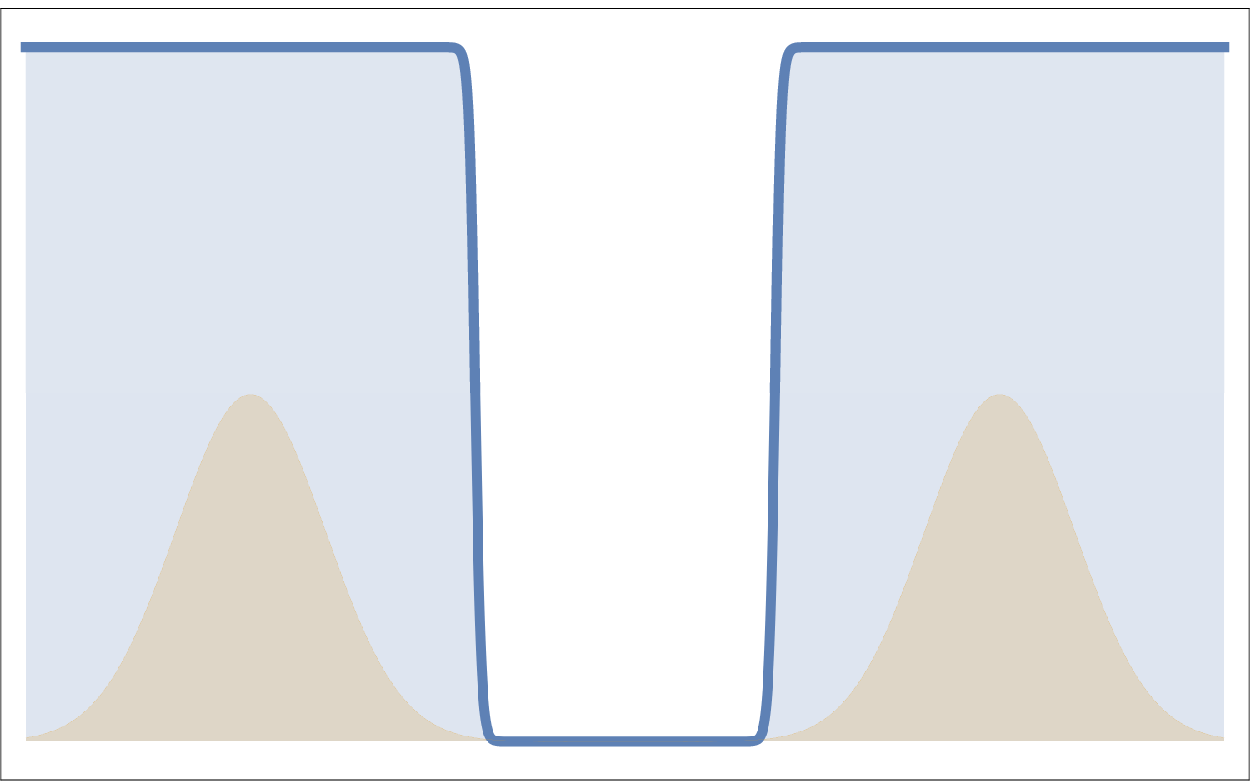}}
\put(43, 70){\includegraphics[width=4.5 cm]{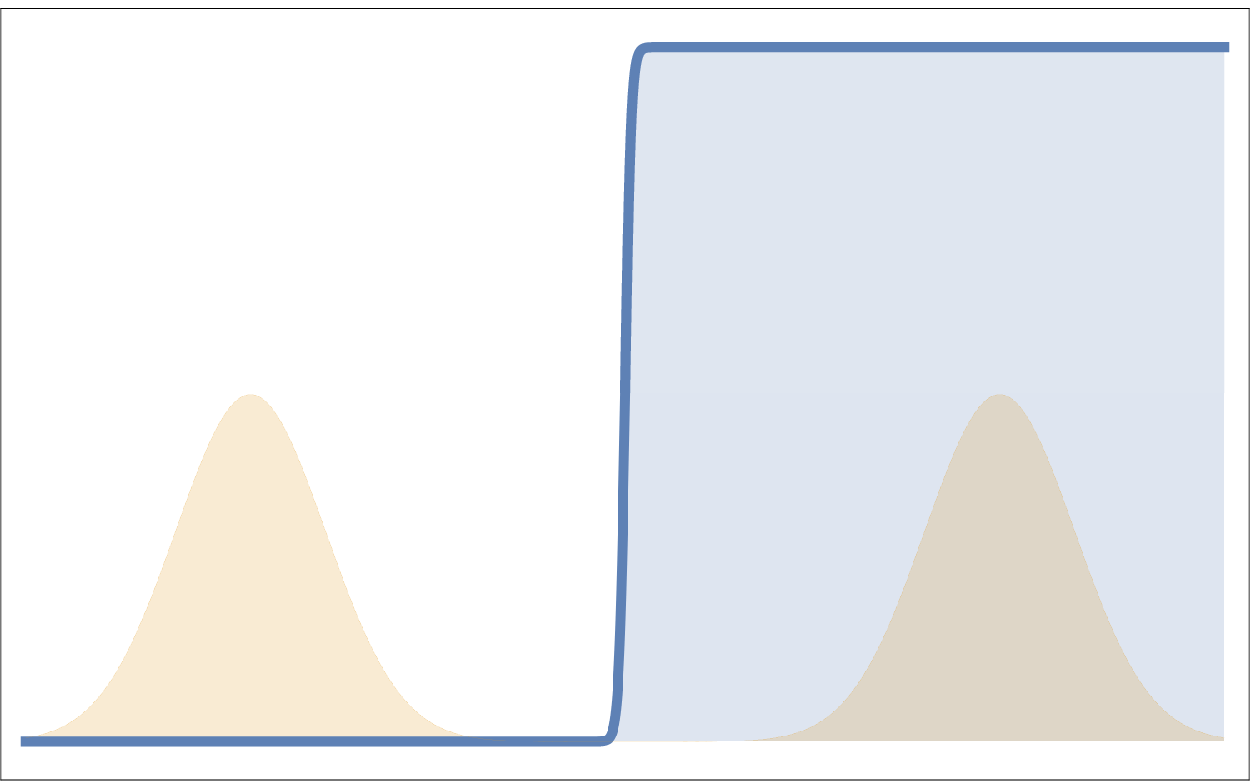}}
\put(96, 70){\includegraphics[width=4.5  cm]{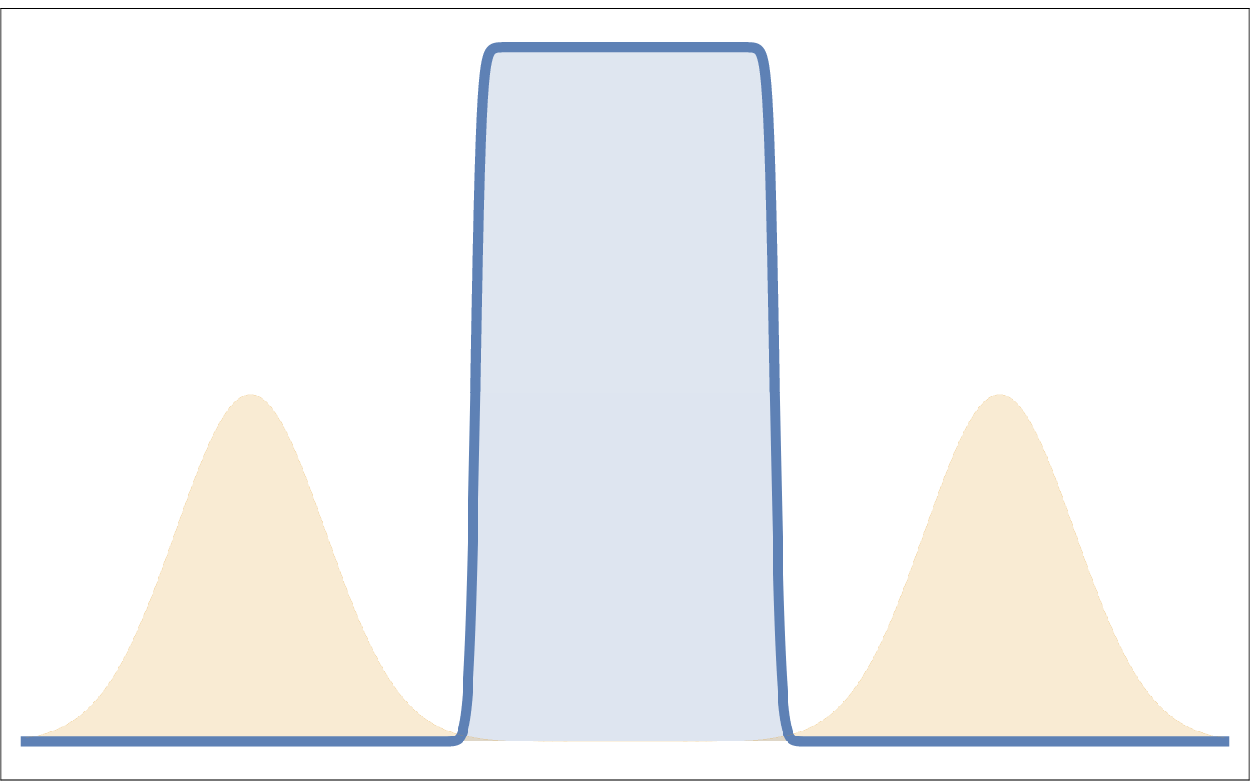}}

\put(-10, 37){\includegraphics[width=4.5 cm]{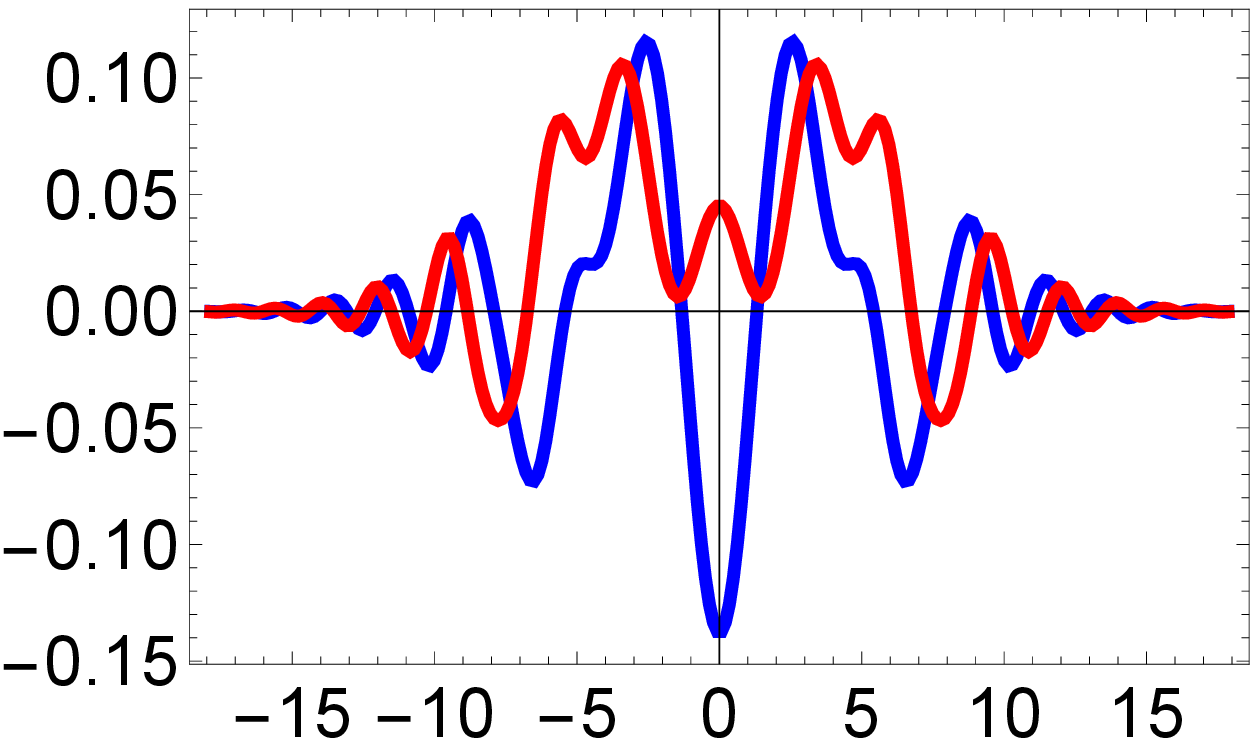}}

\put(42, 37){\includegraphics[width=4.5 cm]{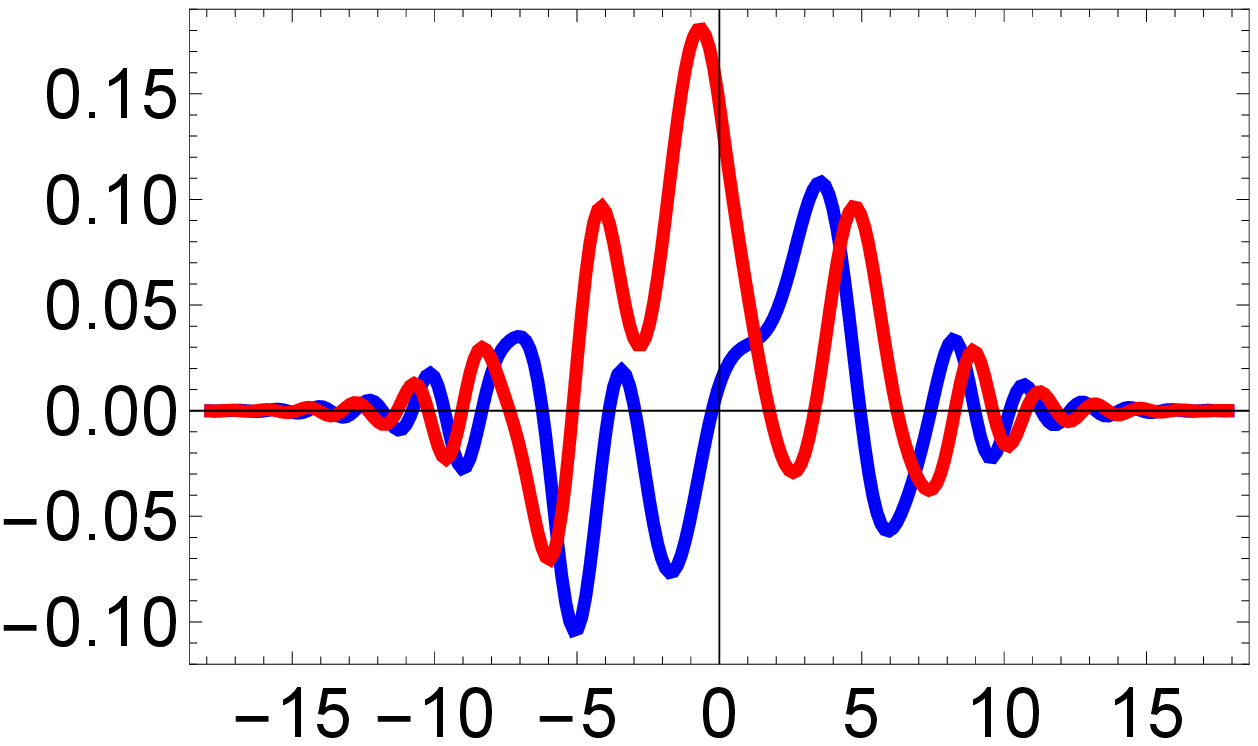}}

\put(95, 37){\includegraphics[width=4.5 cm]{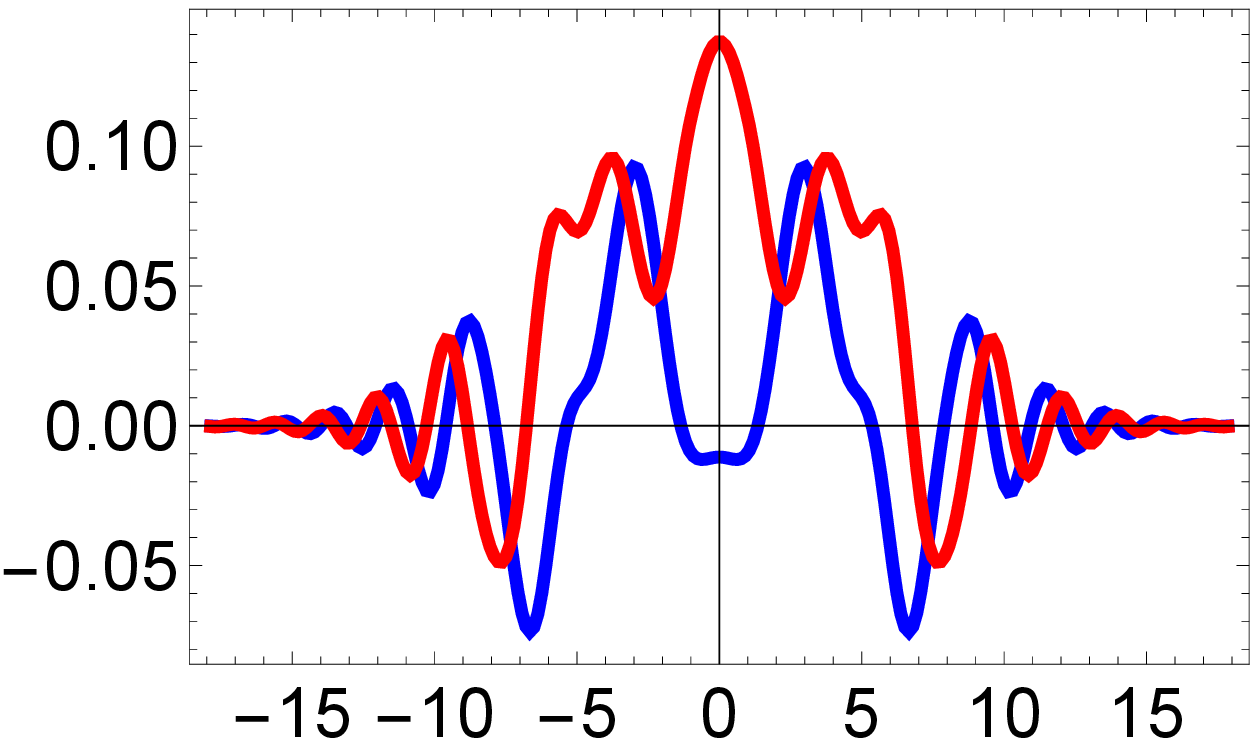}}

\put(-10, 0){\includegraphics[width=4.5 cm]{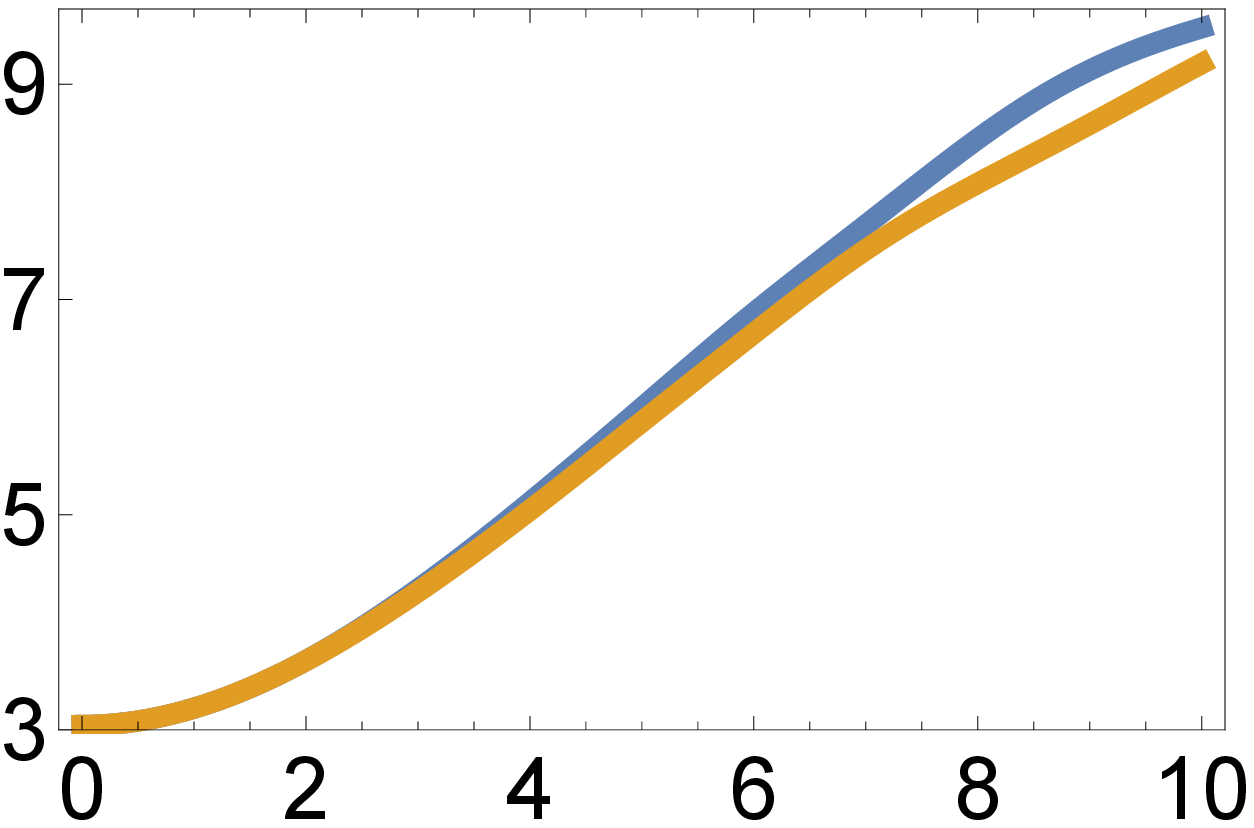}}
\put(42, 0){\includegraphics[width=4.5 cm]{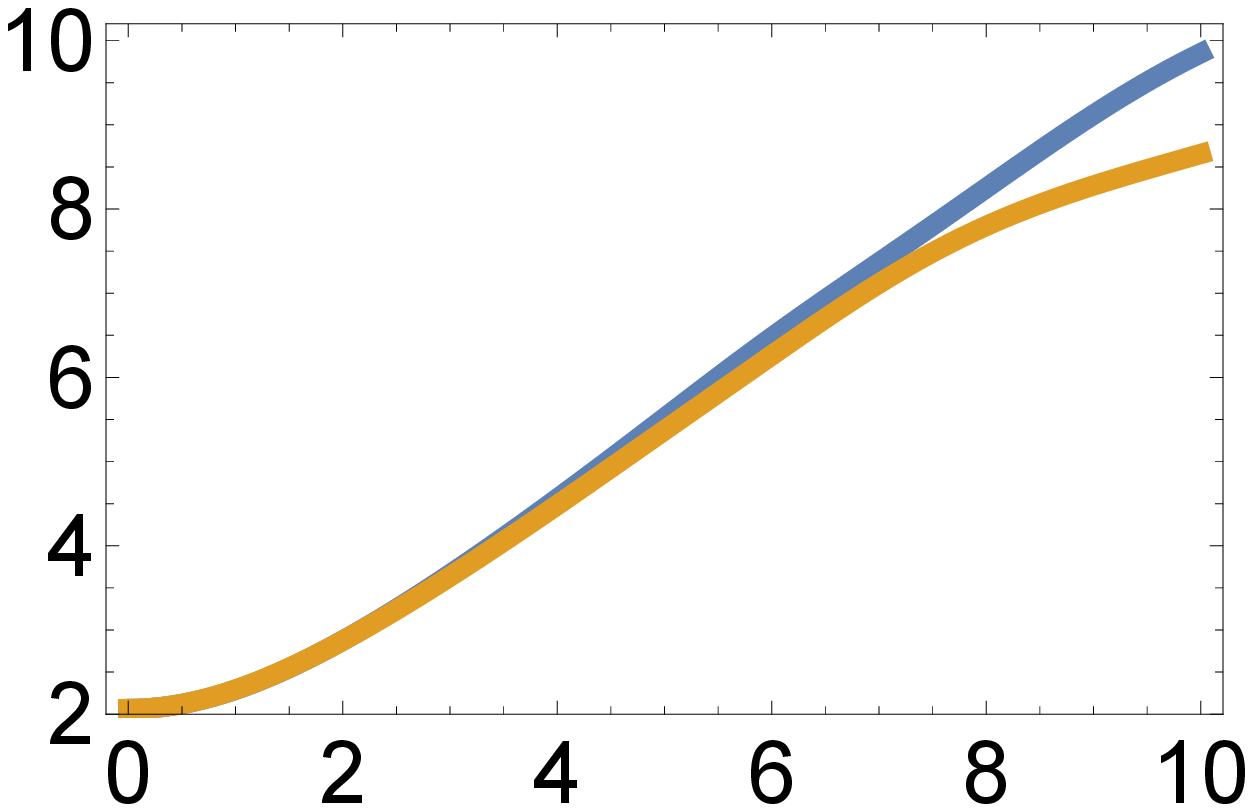}}
\put(95, 0){\includegraphics[width=4.5 cm]{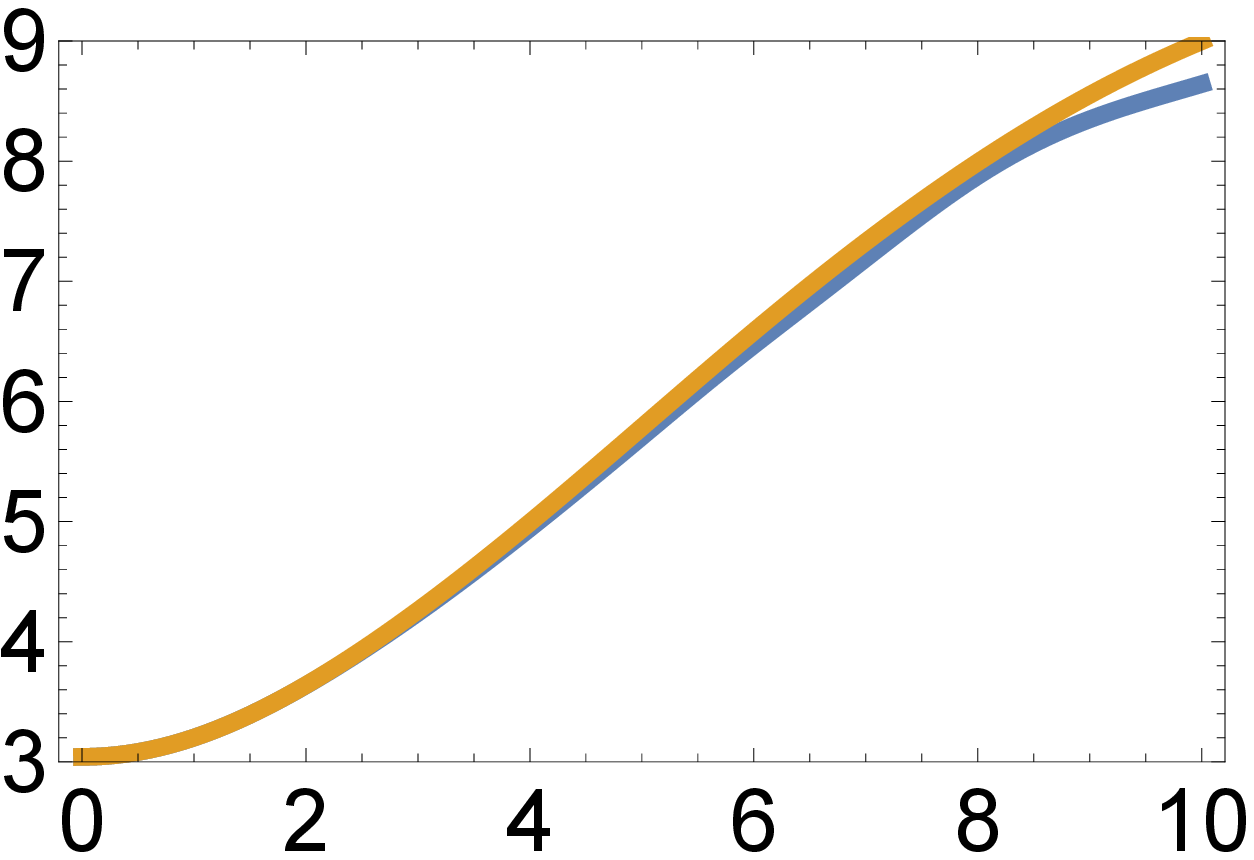}}

\put(-4.5, 16.5){\includegraphics[width=1.8 cm]{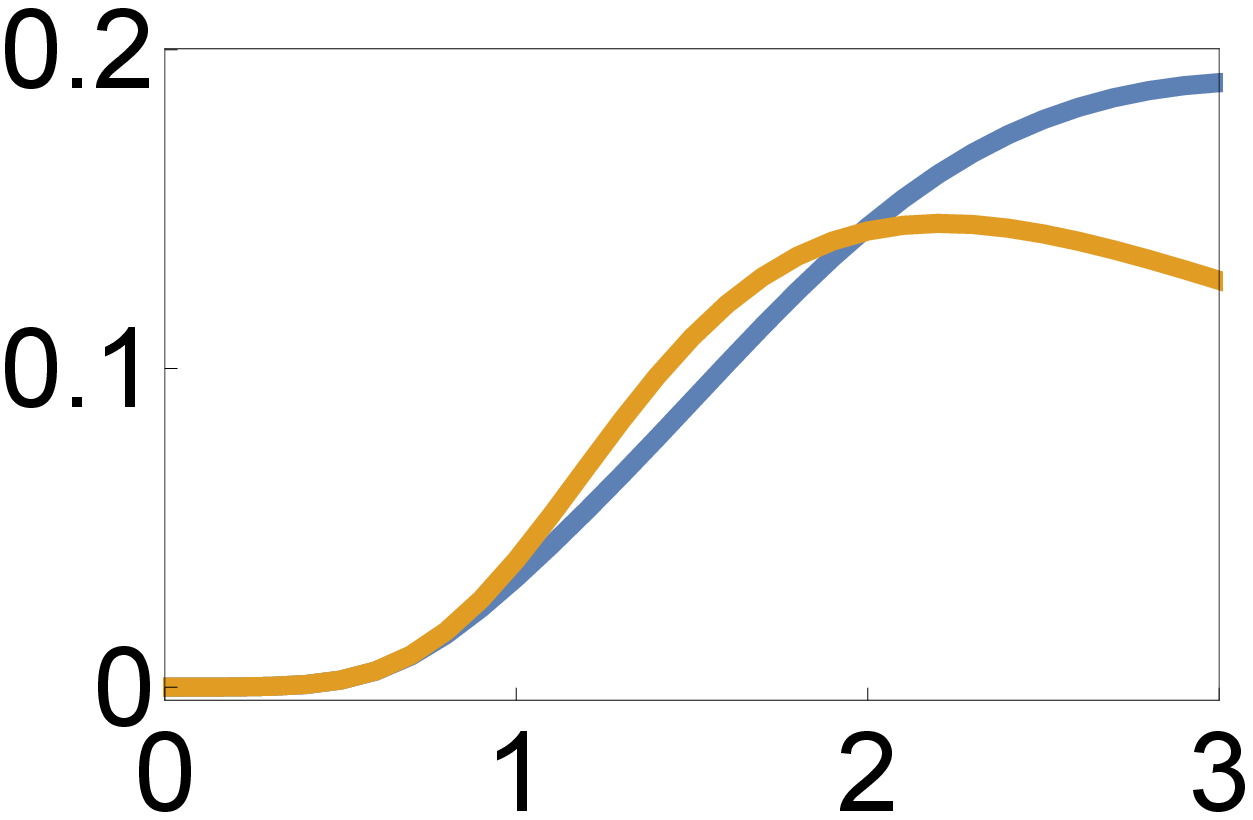}}
\put(49, 16){\includegraphics[width=1.8 cm]{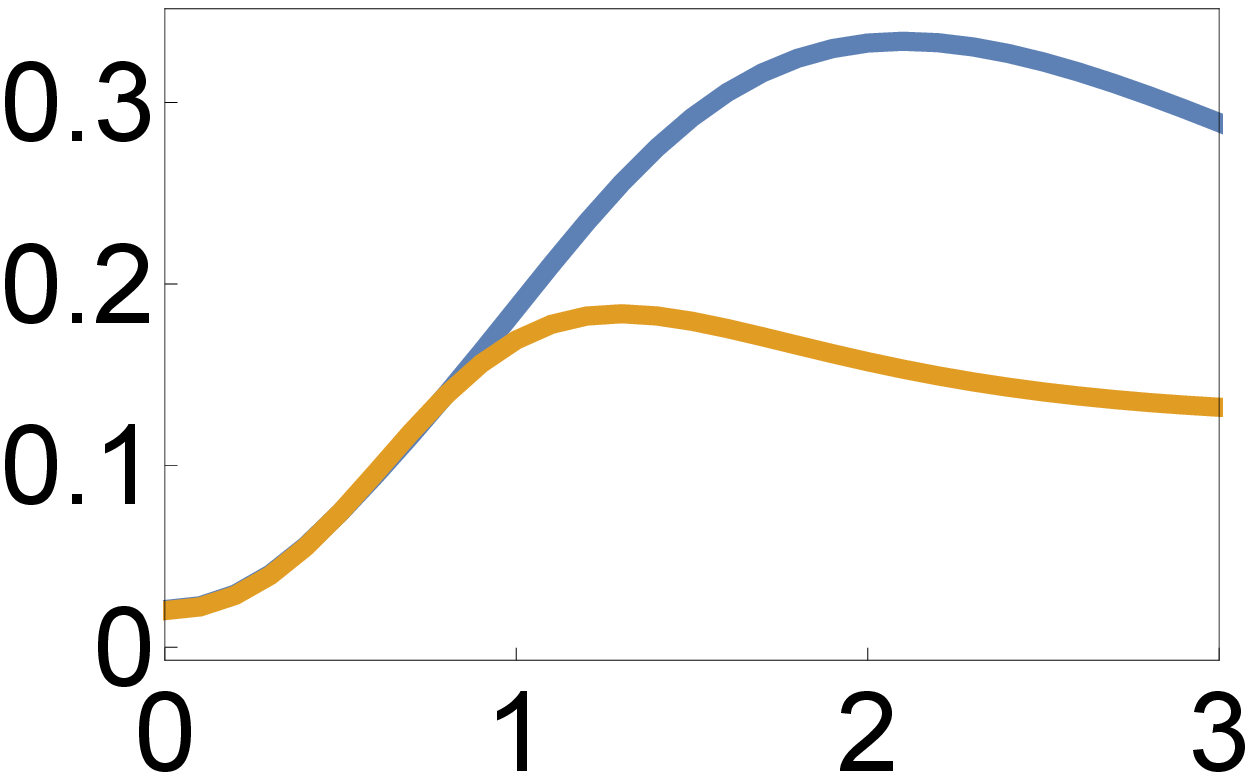}}

\put(101, 17.5){\includegraphics[width=1.8 cm]{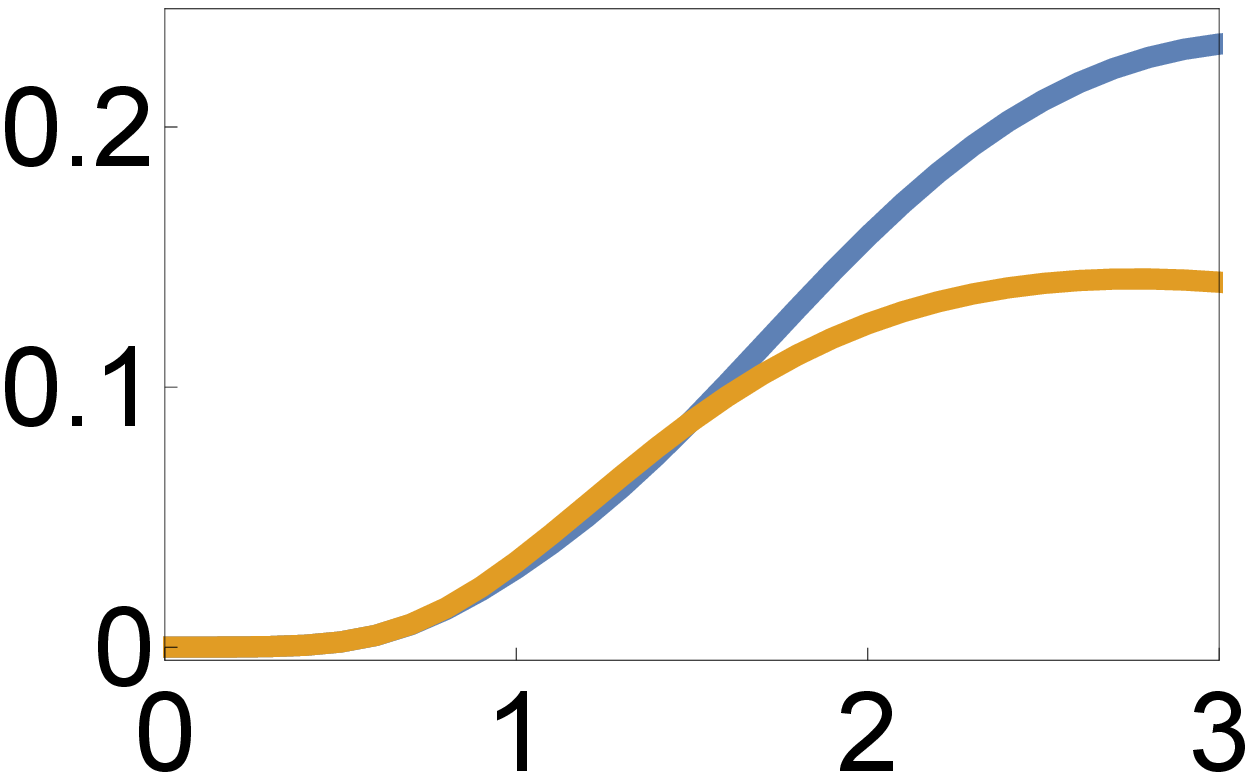}}


\scriptsize
\put(-12, 100){\textbf{a}}
\put(40, 100){\textbf{b}}
\put(93, 100){\textbf{c}}

\put(-12, 65){\textbf{d}}
\put(40, 65){\textbf{e}}
\put(93, 65){\textbf{f}}

\put(-12, 31){\textbf{g}}
\put(40, 31){\textbf{h}}
\put(93, 31){\textbf{i}}

\put(13, 34.5){$x$ (m)}
\put(63, 34.5){$x$ (m)}
\put(118, 34.5){$x$ (m)}
\put(-14, 43) {\rotatebox{90} {Re($\phi$), Im($\phi$) } }

\put(12, -3){$t$ (s)}
\put(63, -3){$t$ (s)}
\put(118, -3){$t$ (s)}

\put(-15, 10) {\rotatebox{90} {$\Delta x^2$ (m$^2$) } }
\put(38, 10) {\rotatebox{90} {$\Delta x^2$ (m$^2$) } }
\put(90, 10) {\rotatebox{90} {$\Delta x^2$ (m$^2$) } }

\tiny
\put(-6.5, 22) {\rotatebox{90} {$\mu$ } }
\put(2.5, 14.5){$t$ (s)}

\put(47, 22) {\rotatebox{90} {$\mu$ } }
\put(56, 14.5){$t$ (s)}

\put(98.5, 23) {\rotatebox{90} {$\mu$ } }
\put(108, 15.5){$t$ (s)}

}
\end{picture}
\caption{\textbf{Transport of quantum matter with different amounts of correlations across different boundaries.} \textbf{a-c} Fragments of quantum matter (yellow) initially localized on the left and right hand sides of a square potential well (\textbf{a}), a potential step (\textbf{b}) and a square potential barrier (\textbf{c}). \textbf{d-f} The real (blue) and imaginary (red) parts of the order parameter ($\phi$) of $\Psi_{\text{SF}_1}$ at $t=3$ s for the three potential structures in \textbf{a, b} and \textbf{c} respectively. \textbf{g-i} The growth of the position variance $\Delta x^2$ for the three potential structures respectively for the initial states $\Psi_{\text{SF}_1}$ (blue) and $\Psi_{\text{MI}}$ (orange). The insets in \textbf{g}-\textbf{i} depict the growth of the fraction of particles $\mu$ trapped inside the region $|x|<1$ which corresponds to the boundaries of the well and barrier in \textbf{g} and \textbf{i}.
   }
\label{fig2} 
\end{figure*}

The position variance $\Delta x^2$ for both states as they evolve in time is depicted in Figs. 2g-2i in the three cases. We notice that for the barrier potential, the variance of the MI state grows slightly faster than in the SF state,  as opposed to the case for the well potential. We also depict in the insets of Figs. 2g and 2i the fraction $\mu$ of the population that accumulate inside the well and the barrier potential respectively. We notice that initially, the particles in the SF state are transported inside the barrier potential faster than the particles in the MI state, whereas the opposite takes place in the well potential, in consistency with the behavior of $\Delta x^2$. We conclude that in the former case, the coherence assisted the tunneling of atoms, while in the later case it has suppressed the initial transport of atoms. The distinction between the two cases can be understood in view of the change of the phase of the order parameter of the SF state after crossing the potential boundaries in both cases. Since the SF atoms are non-interacting,  their behavior is described solely by the evolution of $\phi_0(x)$ in a single particle system. We show in Figs. 2d and 2f snapshots of the real and imaginary parts of $\phi_0(x)$ at $t=3$ s, from which we can anticipate the correlation or anti-correlation between the fragment transported to the central region ($|x|<1$) and the outer fragments (consider the imaginary part of $\phi_0(x)$ for the well potential and the real part of  $\phi_0(x)$ for the potential step). This (anti-)correlation  gives rise to attractive or repulsive forces between the fragments in the middle and the outer fragments in agreement with the viewpoint presented above.

The step potential at $x=0$ in Fig. 2b is more complicated; it can be regarded as a barrier potential for one fragment that tunnels through it, and a potential well for the other fragment on the other side. We can notice from the plots of $\phi_0(x)$ at $t=3$ s in Fig. 2e the anti-correlation between the left and right sides for the real part and the correlation for the imaginary parts. The calculation of $\Delta x^2$ as shown in Fig. 2h implies an overall more repulsion for the SF state (which is an anti-correlation effect) while the calculation of the fraction of atoms accumulating in the region $x<|1|$ as shown in the inset of Fig. 2h indicates a more attractive force for the SF state. The correlation functions for each of the two states in all the three cases are given in the Supplementary Material.


\begin{figure*}[t!] \setlength{\unitlength}{0.1cm}
\begin{picture}(90 , 65 )
{

\put(-10, 49){\includegraphics[width=5 cm ]{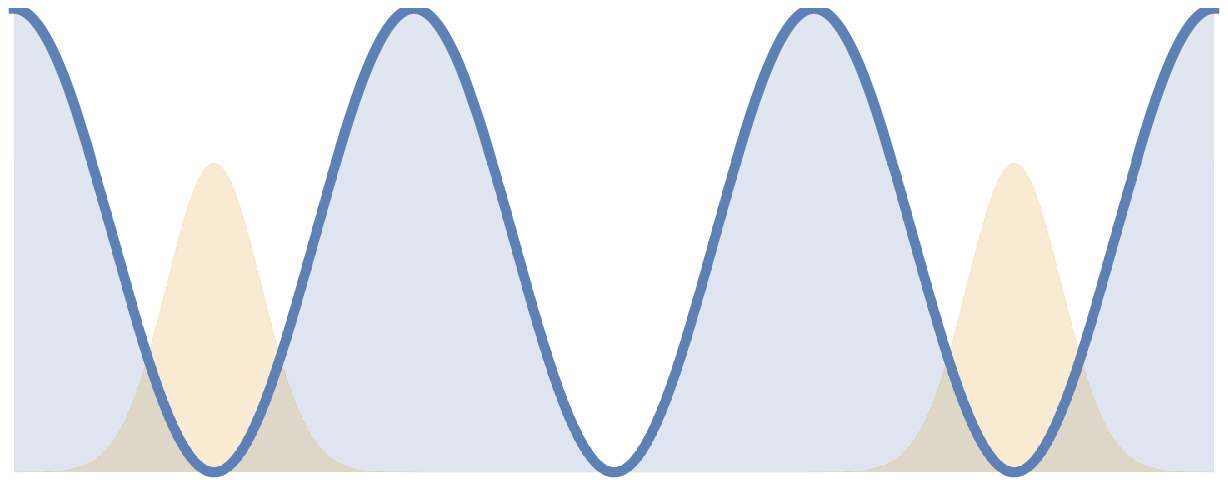}}
\put(49, 39){\includegraphics[width=5 cm]{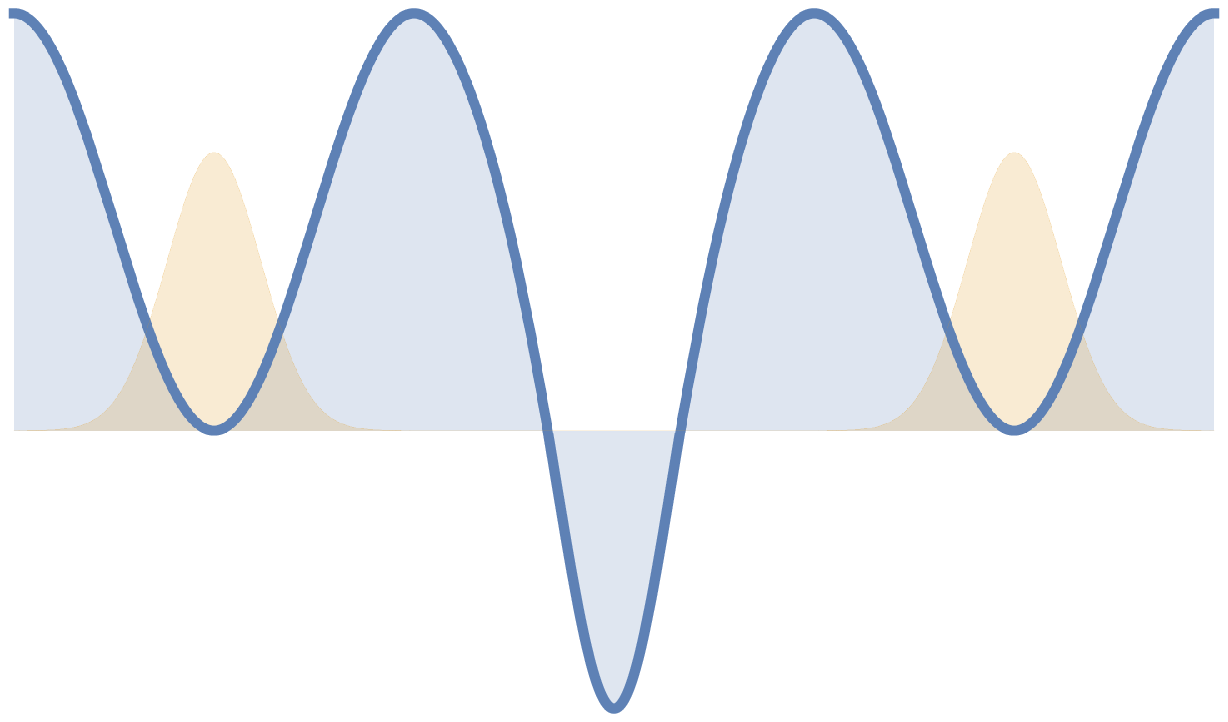}}

\put(-9, 4){\includegraphics[width=5 cm]{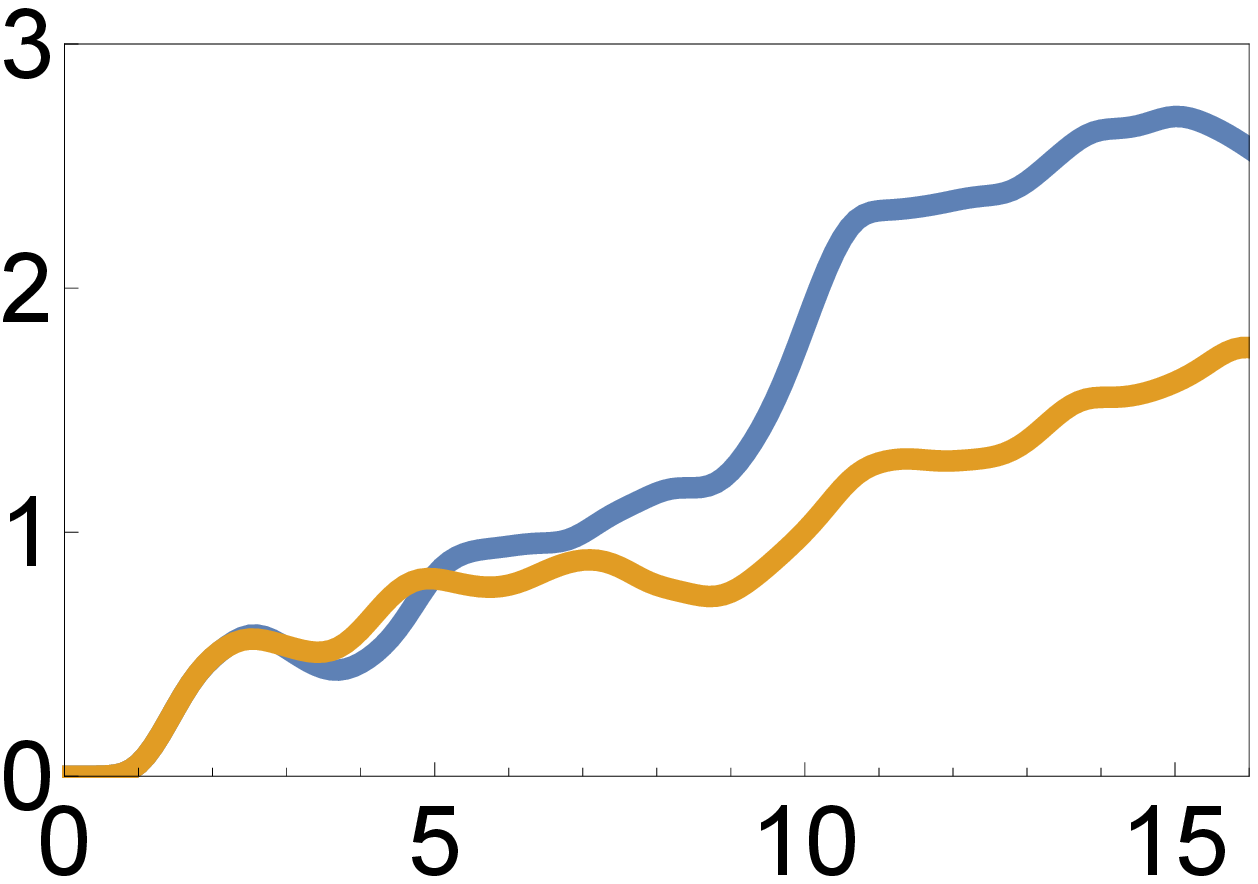}}
\put(49, 4){\includegraphics[width=5 cm]{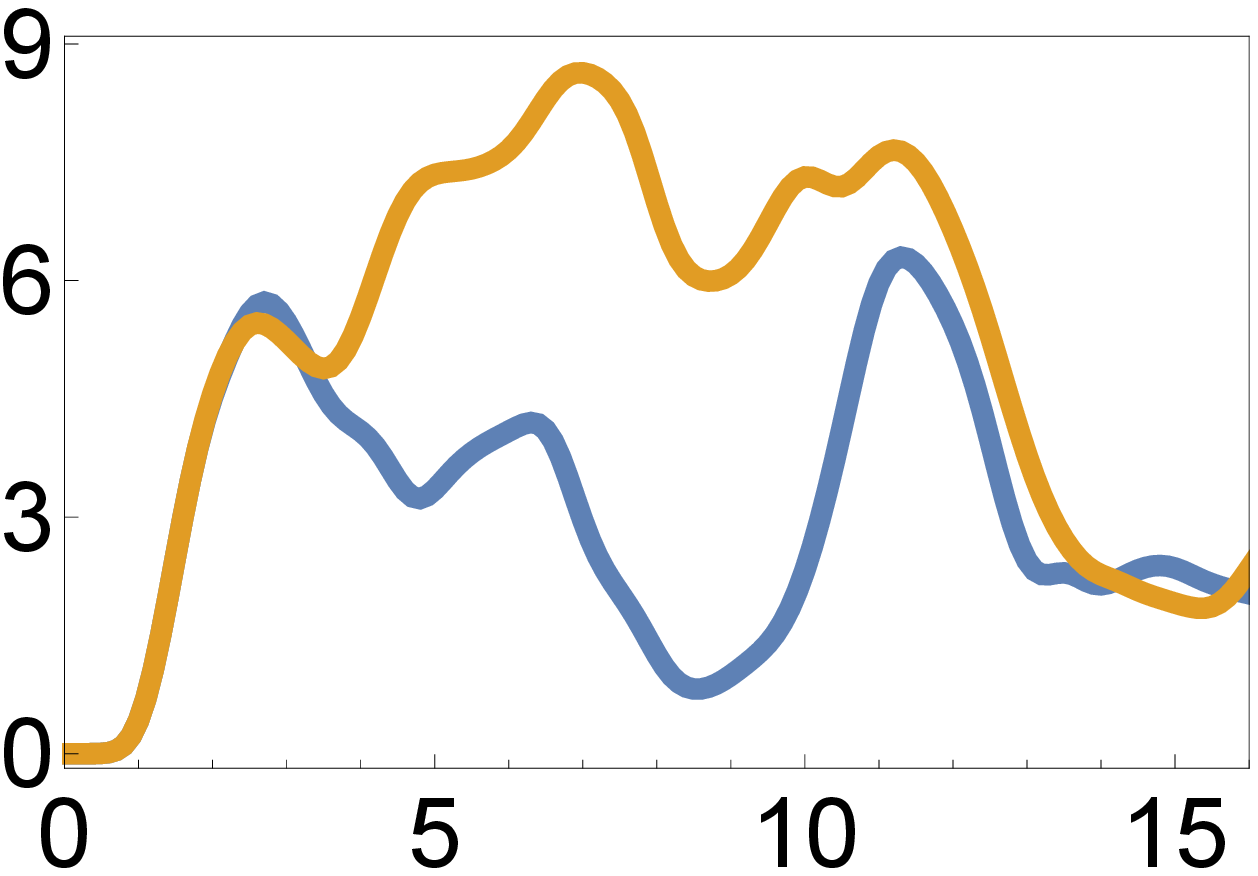}}

\put(14, 0){$t$ (s)}
\put(-15, 22) {\rotatebox{90} {$\mu$ } }
\put(74, 0){$t$ (s)}
\put(44, 22) {\rotatebox{90} {$\mu$ } }

\scriptsize

\put(-13, 69){\textbf{a}}
\put(45, 69){\textbf{b}}
\put(-13, 42){\textbf{c}}
\put(45, 42){\textbf{d}}

\put(-9, 40){$\times 10^{-2}$}
\put(49, 40){$\times 10^{-3}$}
}
\end{picture}
\caption{\textbf{Tunneling of correlated and uncorrelated fragments of quantum matter in a triple-well potential.} \textbf{a,b} Fragments of quantum matter initially localized at the left and right wells of a triple-well potential without a dip  and with a dip in the middle respectively. \textbf{c,d} The growth of the trapped fraction of particles for the initial states $\Psi_{\text{SF}_1}$ (blue) and $\Psi_{\text{MI}}$ (orange) in the middle well for  \textbf{a} and \textbf{b} respectively.  }
\label{fig3} 
\end{figure*}

\section{Applications to tunneling experiments}
We can notice the effect of the coherence-induced forces in a typical tunneling experiment of cold atoms in an optical lattice. We give an example in Fig. 3, where two quantum fragments are trapped in the leftmost and rightmost wells of a triple-well potential (TWP) of the form $V(x)=\alpha \sin^2(kx) $  and are allowed to tunnel through to the middle well. We also simulate a variation of this experiment where a gaussian dip is added to the middle well to reproduce the effect of the repulsive quantum forces  in Fig. 2a. The growth of the fraction of atoms $\mu$ that makes it to the middle well with time in the two cases is plotted for both $\Psi_{\text{SF}_1}$ and $\Psi_{\text{MI}}$ in Figs. 3c and 3d respectively. The triple-well potential in Fig. 3a is given by $V(x)=3\sin^2(\frac{\pi}{6} x)$ and in Fig. 3b, a dip of gaussian shape $w(x)=-2e^{-x^2}$ is added to $V(x)$. The fraction of atoms $\mu$ plotted in Figs. 3c and 3d corresponds to the atoms trapped in the region $|x|<3$. 

Again, we can see the clear distinction between the behavior of $\Psi_{\text{SF}_1}$ compared to $\Psi_{\text{MI}}$ in the two cases. The SF atoms that tunnel through to the middle well are correlated with the two main fragments on the left and the right while the MI atoms that tunnel through to the middle well are at most correlated with one fragment only on the left or the right wells, thus inducing less forces than the former case. In the TWP without a dip, atoms tunneling through to the middle well are in phase with their original fragments, thus inducing attractive forces that slightly accelerates the tunneling, while the SF atoms that tunnel to the middle well with a dip in Fig. 3b acquire an opposite phase that induces repulsive forces that suppress the tunneling.

Note that if each of the two systems considered in Fig. 3 consisted of one single fragment in one well only, the tunneling amount to the middle well will still be different for the TWP with a dip and without a dip for the same reason. We provide the results of this experiment in Fig. 4 where we show the tunneling of a single fragment initially localized at the right well in a triple-well potential for the two cases in Fig. 3: without a dip in the middle and with a dip. The different tunneling behaviors are due to the different interference between the tunneling fragment and the original fragment in the two cases as explained above.

\begin{figure*}[h!] \setlength{\unitlength}{0.1cm}
\begin{picture}(80 , 55 )
{

\put(-15, 5){\includegraphics[width=5 cm ]{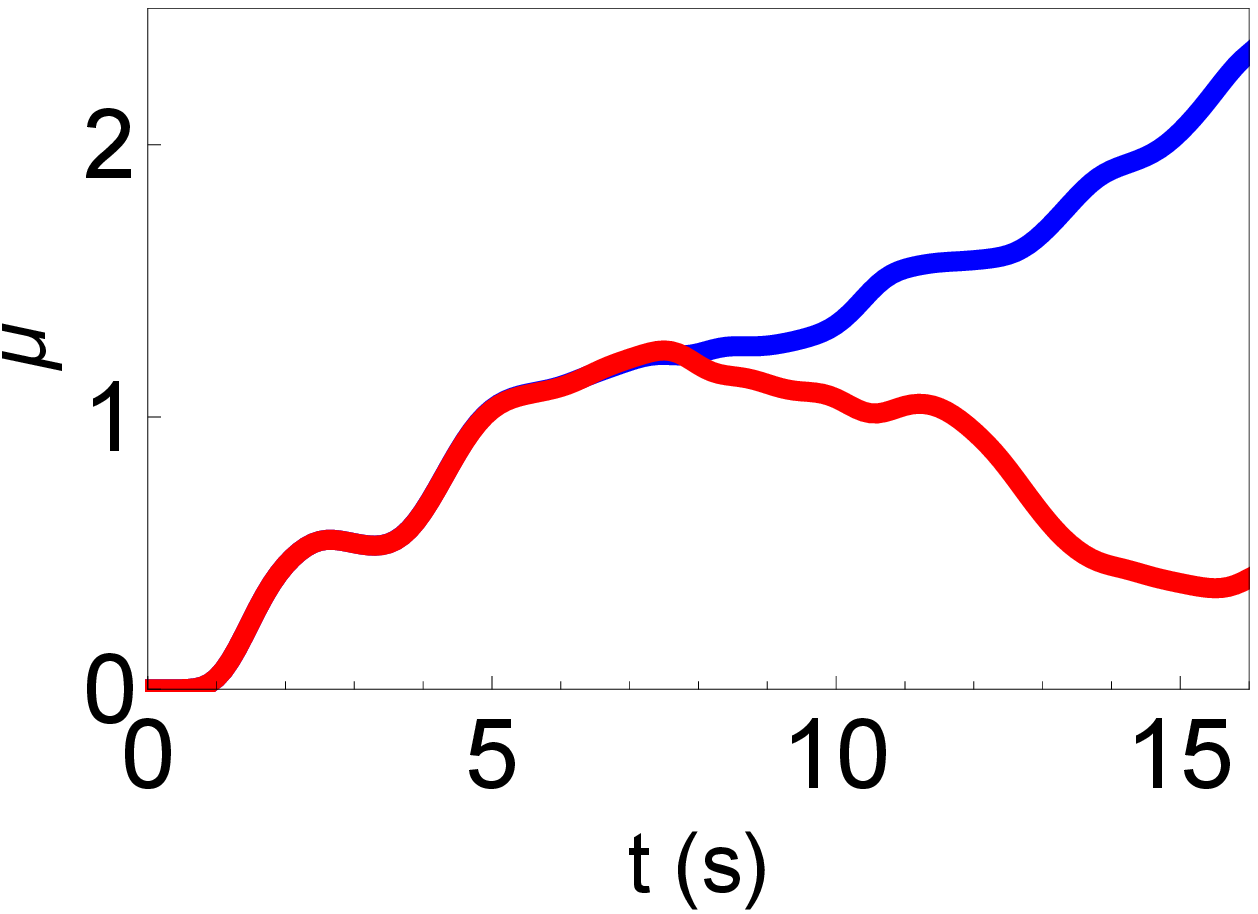}}
\put(50, 30){\includegraphics[width=4 cm]{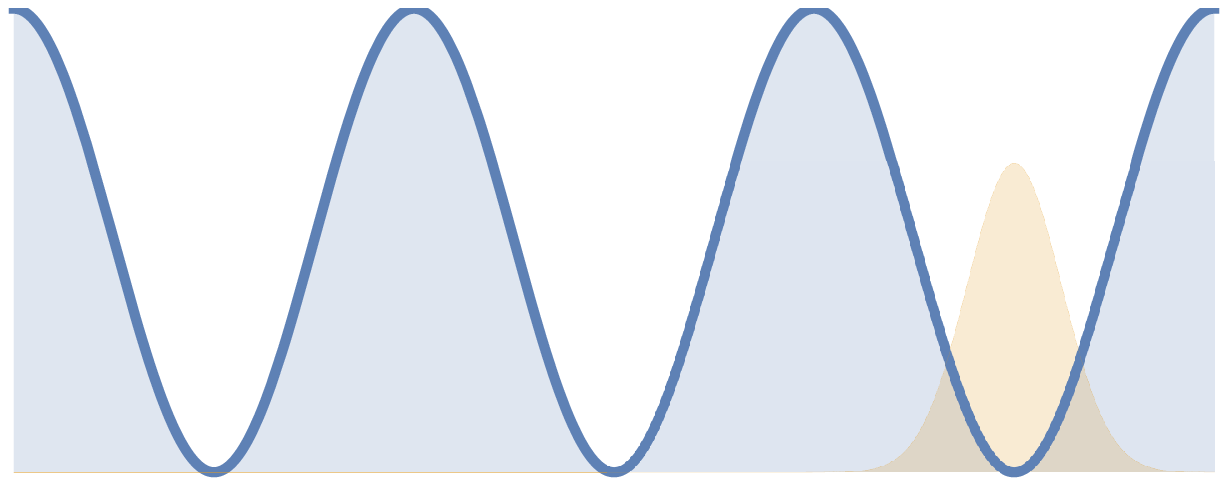}}
\put(50, 0){\includegraphics[width=4 cm]{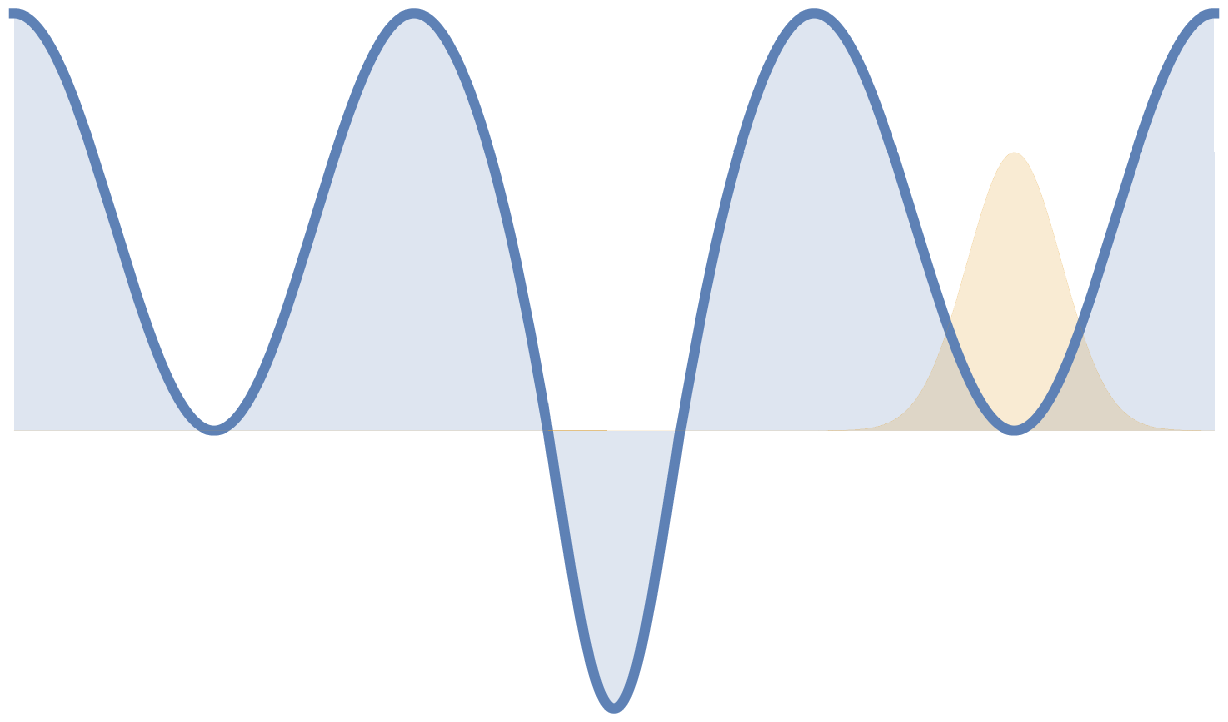}}

\scriptsize
\put(-17, 43){\textbf{a}}
\put(47, 50){\textbf{b}}
\put(47, 25){\textbf{c}}

\put(-11, 43){$\times 10^{-2}$}

}
\end{picture}
\caption{\textbf{The tunneling of a single fragment in a triple well potential.} The growth of the fraction of atoms $\mu$ tunneling to the middle and left wells for the the triple-well potential without a dip (blue) and with a dip (red). }
\label{fig-4}
\end{figure*}

The effects of the coherence on the tunneling of atoms can also be observed in experiments of shaken optical lattices constructed by driven periodic potentials \cite{eckardt2017,zenesini2010}. We give such an example in Fig. 5, where we spatially modulate an optical lattice by periodically shaking the potential back and forth at a certain frequency. Again, we consider three states with different correlations;   $\Psi_{\text{MI}}$ where the localized fragments have no first order correlations,  $\Psi_{\text{SF}_1}$ where the fragments form a superfluid with a strong correlation between nearby sites and  $\Psi_{\text{SF}_2}$ where the SF fragments exhibit an  anti-correlation between nearby sites. The periodically driven lattice used in this example has the potential function $V(x,t)=\frac{72}{\pi^2}\sin^2(\frac{\pi}{6} x+\frac{1}{4}\sin(30t))$. That particular choice of the height of the lattice ensures that the initial gaussian orbitals are close to the ground states of their respective potential wells up to a second order approximation. $\mu$ represents the fraction of atoms trapped in a region within a distance $\delta x\leq 1$ from the peaks of $V(x,0)$. We plot the fraction of matter that accumulate between the wells  (at the crests of the potential function of the undriven lattice) during a few cycles for the three states in Fig. 5a. As can be readily noticed, the tunneling is enhanced (suppressed) for  $\Psi_{\text{SF}_1}$($\Psi_{\text{SF}_2}$) compared to $\Psi_{\text{MI}}$.


\begin{figure*}[t!] \setlength{\unitlength}{0.1cm}
\begin{picture}(90 , 37 )
{

\put(50, 30){\includegraphics[width=5 cm ]{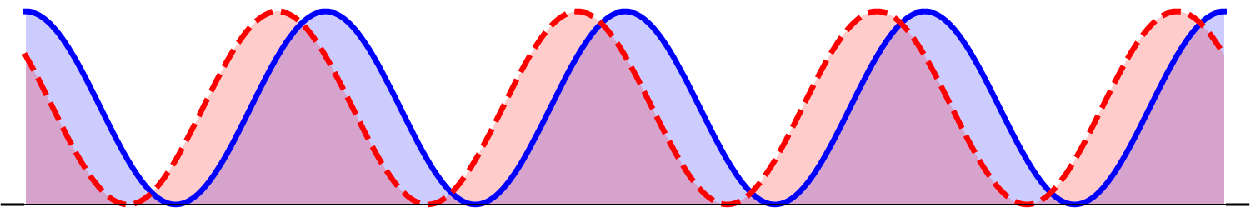}}
\put(50, 18){\includegraphics[width=5 cm]{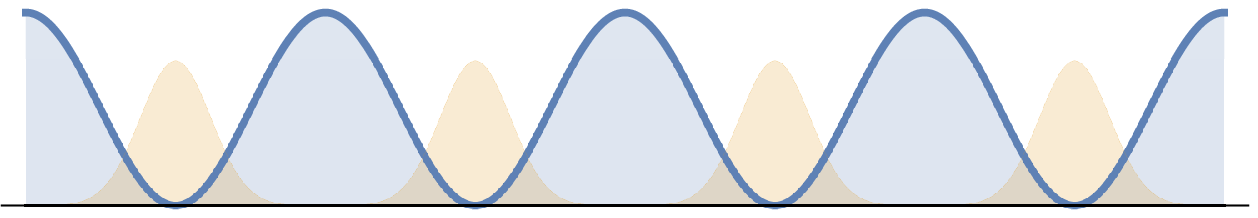}}
\put(50, 0){\includegraphics[width=5 cm]{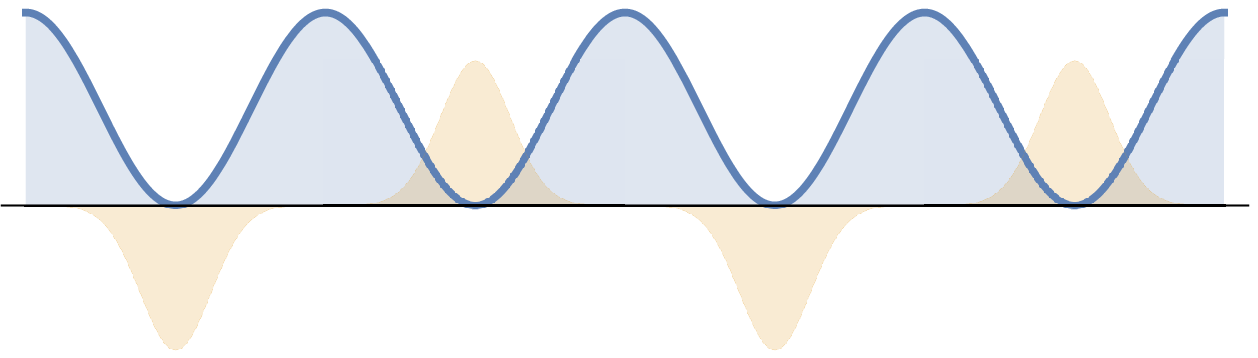}}
\put(-15, 5){\includegraphics[width=6. cm]{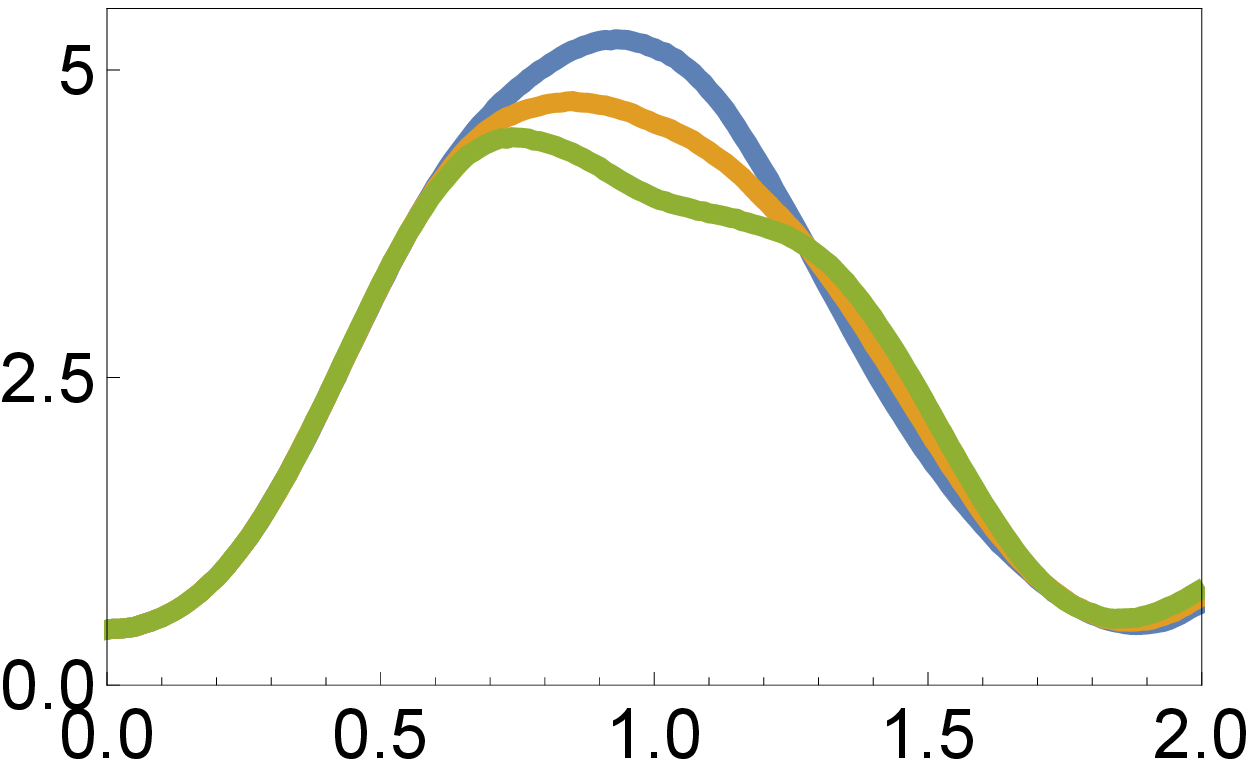}}

\put(57, 40){\Arrow}
\put(69, 40){\Arrow}
\put(81, 40){\Arrow}

\put(17, 0){$t$ (s)}
\put(-19, 22) {\rotatebox{90} {$\mu$ } }

\scriptsize
\put(-17, 42){\textbf{a}}
\put(47, 37){\textbf{b}}
\put(47, 25){\textbf{c}}
\put(47, 12){\textbf{d}}

\put(-13, 42){$\times 10^{-4}$}

}
\end{picture}
\caption{\textbf{Tunneling of localized fragments of quantum matter with different correlations in a periodically driven optical lattice.} \textbf{a} The growth of the fraction of trapped atoms $\mu$ inside the locations of the crests of the undriven optical lattice (\textbf{b}) for the initial states  $\Psi_{\text{SF}_1}$ (blue), $\Psi_{\text{MI}}$ (orange) and $\Psi_{\text{SF}_2}$ (green). \textbf{c,d} The initial orbitals for the states $\Psi_{\text{SF}_1}$ and $\Psi_{\text{SF}_2}$ respectively. }
\label{fig5}
\end{figure*}

\section{Discussion}
We have shown, based on ab-initio calculations, that the existence of quantum coherence between nearby fragments of quantum matter of non-interacting particles leads to the onset of an effective force between the particles that is absent for uncorrelated fragments. This force  is best manifested in the transient behavior of  the transport dynamics across potential barriers, which can lead to either assisted or suppressed tunneling based on the sign of correlation. We can attribute this force to the interference between the coherent fragments, which contribute into the total Bohmian force felt by each quantum particle. The direction of this force depends on the sign of the correlation function; attractive for positive correlation and repulsive for negative correlation. It is very likely that the effect of this force is more noticeable in situations where the fraction of tunneling particles is so tiny that a strong correlation between the tunneling particles and the originally uncorrelated fragments does not develop. The predictive power of this model is limited only by our ability to quantify the amount of interference and predict whether constructive or destructive interference will prevail. This becomes more complicated when the fragments start propagating.
For freely propagating correlated fragments that incur varying phase factors extending over large regions due to their propagation, both constructive and destructive interference will take place and they will have opposite effects.

We introduced two proof-of-concept situations in multi-well traps and optical lattices where the distinction between the transport of correlated and uncorrelated fragments  can be observed experimentally.  Our simulation focused on non-interacting particles only without any loss of generality and the same principle is applicable for interacting quantum particles. We should bear in mind that the framework of relating the transport of quantum matter to the coherence properties presented here is in direct correspondence with the proposal that noise assisted tunneling in biological systems works by  causing the loss of destructive interference \cite{caruso2009}. While our numerical experiments indicate a direct proportionality between the transport and the amount of coherence, quantifying this effect in a rigorous manner still needs to be done. It is clear that all the dynamics presented in this paper can be understood from first principles without resorting to assuming the existence of effective forces. However, introducing this concept will help explain different tunneling phenomena in terms of the coherence properties directly in an intuitive manner without going down to the detailed dynamics of the many-particle wavefunction.

There are states, other than the MI state, that exhibit a lack of phase correlation between different sites and can be contrasted with the SF state in order to observe the effect of the coherence-induced forces. One example is the N00N state \cite{wildfeuer2007} defined as $\frac{1}{\sqrt{2}}\left( |N\rangle_L |0\rangle_R +e^{i\phi}|0\rangle_L |N\rangle_R \right) $, where $|n\rangle_L$ and $|n\rangle_R$  are  number states representing the occupancies of  two orthogonal orbitals corresponding to  the left and right fragments as in Fig. 3.  Another example is a simple statistical mixtures between the left and right fragments that may be caused by the decoherence process. States with partial coherence are expected to exhibit intermediate behavior between the fully correlated SF state and the fully uncorrelated states.

The concept of a quantum force of pure quantum origin introduced in this article may hint at the possibility of the emergence of a similar force of a quantum nature on a macroscopic level. While finding such a quantum origin for the gravitational force for example is a theoretical physicist's dream, it is clear that the effective force introduced above differs in nature from the gravitational force. Nevertheless, it remains to be investigated whether higher order correlations between nearby fragments may lead to some other types of effective forces.

It is very plausible that the different rates of atom tunneling reported above imply different tunneling times, i.e., the amount of coherence affects the tunneling time as well. This will be an interesting result in its own right that may contribute to the ongoing investigation of the tunneling time problem \cite{hauge1989,sokolovski2018}. Apart from that, a deeper understanding of the reported phenomenon may open the door for new applications that take advantage of the distinctive tunneling behavior of correlated particles across boundaries.

\bibliographystyle{apsrev4-1}
\bibliography{force}

\begin{thebibliography}{26}
\bibitem [{not()}]{note}%
  \BibitemOpen
  \href@noop {} {}\bibinfo {note} {In cold atoms experiments, the first order
  correlation function is usually defined as the absolute of $g^{(1)}$ as
  defined in the text.}\BibitemShut {Stop}%
\end{thebibliography}

\clearpage
\begin{widetext}
 \begin{center}
 \Large{Supplementary Material }
 \end{center}
 \vspace{10 mm}
 \setcounter{figure}{0}
 \renewcommand{\thefigure}{S\arabic{figure}}


\section{Correlation functions for Fig. 2}
 
We show below the real and imaginary parts of the correlation function $g^{(1)}(x_1,x_1')$ for the  states $\Psi_{\text{SF}_1}$, $\Psi_{\text{MI}}$ computed at $t=3$ s for the dynamics in Fig. 2.

\begin{figure*}[h!] \setlength{\unitlength}{0.1cm}
\begin{picture}(90 , 84 )
{

\put(-15, 0){\includegraphics[width=5 cm ]{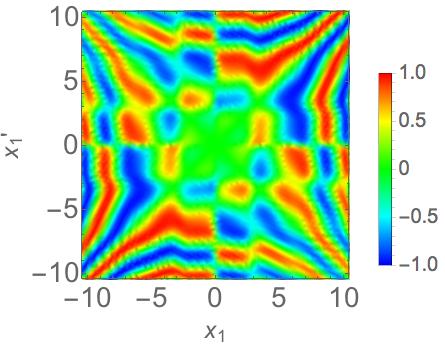}}
\put(50, 0){\includegraphics[width=5 cm]{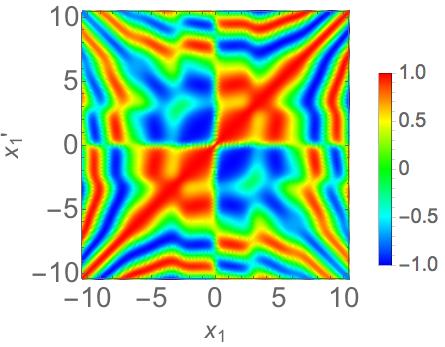}}
\put(-15, 40){\includegraphics[width=5 cm]{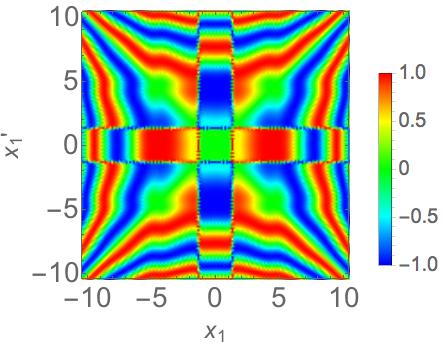}}
\put(50, 40){\includegraphics[width=5. cm]{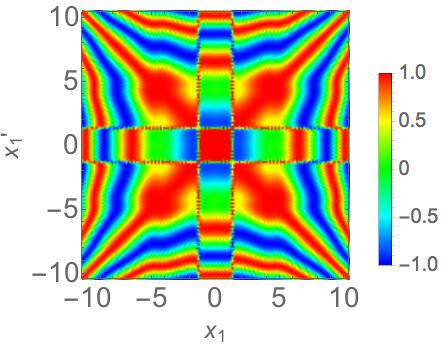}}

\scriptsize
\put(-17, 78){\textbf{a}}
\put(47, 78){\textbf{b}}
\put(-17, 38){\textbf{c}}
\put(47, 38){\textbf{d}}

}
\end{picture}
\caption{\textbf{Correlation functions at $t=3$ s for the well potential in Fig. 2a.} The real (left) and imaginary (right) parts of $g^{(1)}(x_1,x_1')$ for the SF state (a,b) and the MI state (c,d). }
\label{fig-S1}
\end{figure*}

\begin{figure*}[] \setlength{\unitlength}{0.1cm}
\begin{picture}(90 , 84 )
{

\put(-15, 0){\includegraphics[width=5 cm ]{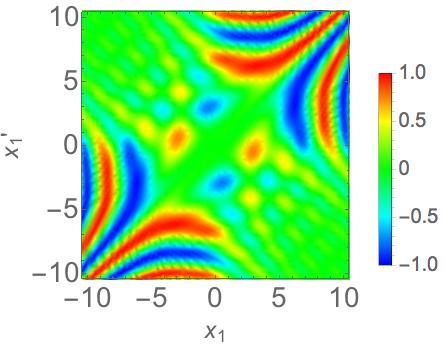}}
\put(50, 0){\includegraphics[width=5 cm]{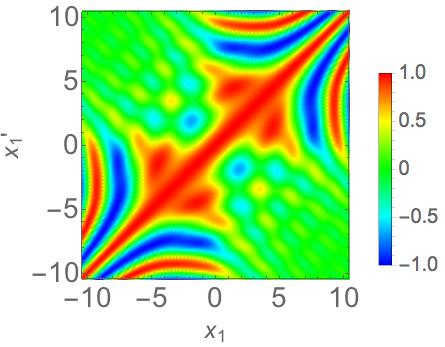}}
\put(-15, 40){\includegraphics[width=5 cm]{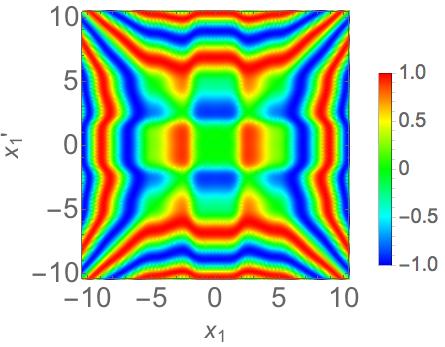}}
\put(50, 40){\includegraphics[width=5. cm]{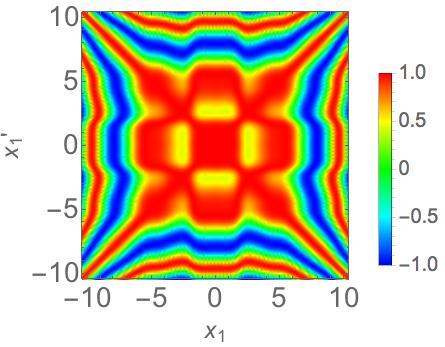}}

\scriptsize
\put(-17, 78){\textbf{a}}
\put(47, 78){\textbf{b}}
\put(-17, 38){\textbf{c}}
\put(47, 38){\textbf{d}}

}
\end{picture}
\caption{\textbf{Correlation functions at $t=3$ s for the step potential in Fig. 2b.} The real (left) and imaginary (right) parts of $g^{(1)}(x_1,x_1')$ for the SF state (a,b) and the MI state (c,d). }
\label{fig-S2}
\end{figure*}

\begin{figure*}[] \setlength{\unitlength}{0.1cm}
\begin{picture}(90 , 84 )
{

\put(-15, 0){\includegraphics[width=5 cm ]{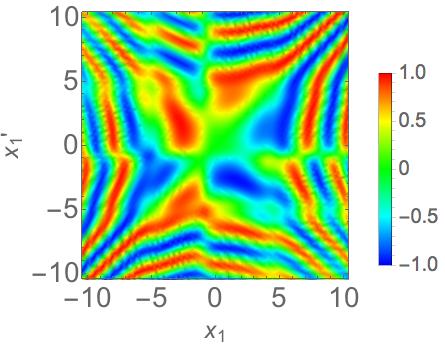}}
\put(50, 0){\includegraphics[width=5 cm]{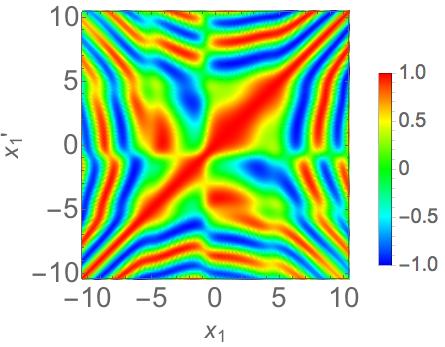}}
\put(-15, 40){\includegraphics[width=5 cm]{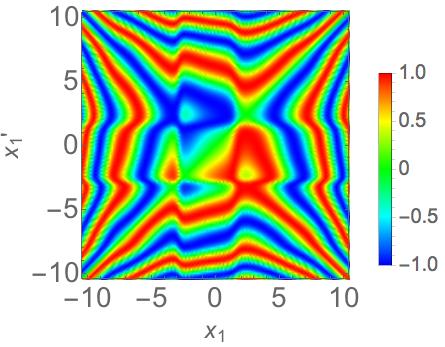}}
\put(50, 40){\includegraphics[width=5. cm]{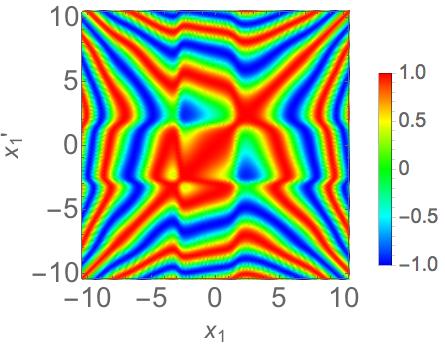}}

\scriptsize
\put(-17, 78){\textbf{a}}
\put(47, 78){\textbf{b}}
\put(-17, 38){\textbf{c}}
\put(47, 38){\textbf{d}}

}
\end{picture}
\caption{\textbf{Correlation functions at $t=3$ s for the barrier potential in Fig. 2c.} The real (left) and imaginary (right) parts of $g^{(1)}(x_1,x_1')$ for the SF state (a,b) and the MI state (c,d). }
\label{fig-S3}
\end{figure*}

\clearpage

\end{widetext}

\end{document}